\documentclass[aps,prb,epsfig,preprint,superscriptaddress,showpacs]{revtex4-2}
\usepackage{epstopdf}
\usepackage{graphicx}  
\usepackage{float}
\usepackage{dcolumn}   
\usepackage{bm}        
\usepackage{amssymb}
\usepackage[english]{babel}
\usepackage{lineno}

\begin{document}

\title{Pressure-driven vibrational and structural peculiarities in the honeycomb layered magnetoelectrics  Mn$_4${\bf \textit{B}}$_2$O$_9$ ({\bf \textit{B}} = Nb, Ta)}

\author{Rajesh Jana}
\email[Contact author: ]{rajesh.jana@hpstar.ac.cn}
\affiliation{Center for High Pressure Science and Technology Advanced Research (HPSTAR), Beijing 100193, P. R. China}
\affiliation{Solid State Physics Division, Bhabha Atomic Research Centre, Mumbai 400085, India}
\author{Afsal S Shajahan}
\affiliation{School of Pure and Applied Physics, Mahatma Gandhi University, Priyadarsini Hills, Kottayam, Kerala 686560, India}
\author{Boby Joseph}
\affiliation{Elettra-Sincrotrone Trieste S.C. p. A., S.S. 14, Km 163.5 in Area Science Park, 34149 Basovizza, Italy}
\author{Brahmananda Chakraborty}
\affiliation{High Pressure and Synchrotron Radiation Physics Division, Bhabha Atomic Research Centre, Mumbai 400085, India}
\affiliation{Homi Bhabha National Institute, Anushaktinagar, Mumbai 400094, India}
\author{Irshad K A}
\affiliation{Elettra-Sincrotrone Trieste S.C. p. A., S.S. 14, Km 163.5 in Area Science Park, 34149 Basovizza, Italy}
\author{Anuj Upadhyay}
\affiliation{Hard X-Ray Applications Lab, Synchrotron Utilisation Section, RRCAT, Indore 452013, India}
\author{Alka Garg}
\affiliation{High Pressure and Synchrotron Radiation Physics Division, Bhabha Atomic Research Centre, Mumbai 400085, India}
\affiliation{Homi Bhabha National Institute, Anushaktinagar, Mumbai 400094, India}
\author{Rekha Rao}
\email[Contact author: ]{rekhar@barc.gov.in}
\affiliation{Solid State Physics Division, Bhabha Atomic Research Centre, Mumbai 400085, India}
\affiliation{Homi Bhabha National Institute, Anushaktinagar, Mumbai 400094, India}
\author{Thomas Meier}
\email[Contact author: ]{thomasmeier@sharps.ac.cn}
\affiliation{Shanghai Key Laboratory MFree, Institute for Shanghai Advanced Research in Physical Sciences, Shanghai 201203, P. R. China.}
\affiliation{Center for High Pressure Science and Technology Advanced Research (HPSTAR), Beijing 100193, P. R. China}

\begin{abstract}
The high-pressure behavior of two Mn-based honeycomb-structured magnetoelectric materials, Mn$_4$Nb$_2$O$_9$ (MNO) and Mn$_4$Ta$_2$O$_9$ (MTO), has been investigated using Raman spectroscopy, synchrotron x-ray diffraction and density functional theory (DFT) calculations. In MTO, application of a small pressure of only 0.5 GPa induces an isostructural transition driven by local symmetry breaking. With further increase in pressure, three additional isostructural transitions are observed, around 3.2, 6, and 10 GPa, followed by the onset of a long-range structural transition near 14 GPa, where the ambient \textit{P-3c1} phase begins to transform to a \textit{P2/c} phase. These two phases coexist up to 27 GPa. The Nb analogue, MNO, also exhibits similar isostructural transitions at approximately 2, 6.6, and 10 GPa; however, the onset of the mixed \textit{P2/c} and \textit{P-3c1} phases occurs at a slightly lower pressure (~12.5 GPa), with coexistence extending up to 26.5 GPa. These long-range transitions are corroborated by pressure-dependent enthalpy variations revealed through DFT computations. Rietveld refinement reveals pronounced anisotropic lattice compression (42–49\%) between the $c$ and $a$ axes, leading to a notable reduction in the \textit{c/a} ratio. This anisotropy may enhance interlayer coupling and promote magnetic ordering under compression, consistent with the emergence of Raman modes resembling those reported at low temperatures, along with anomalous variations in Raman mode linewidth and intensity. The pronounced changes in Raman self energy parameters, anomalies in the reduced pressure–Eulerian strain profile, and the onset of local symmetry breaking at much lower pressures in MTO than in MNO highlight the crucial role of differing spin–orbit coupling strength and orbital hybridization effects associated with Nb$^{5+}$ and Ta$^{5+}$ cations.
\end{abstract}

\maketitle
\section{Introduction}

The honeycomb-layered family of compounds with the general formula $A_4B_2$O$_9$, where divalent transition-metal ions such as Co, Fe, or Mn occupy the $A$-site and pentavalent nonmagnetic ions like Nb or Ta reside at the $B$-site, has recently garnered considerable research attention owing to the emergence of intriguing physical phenomena, including magnetodielectric coupling, spin-driven linear magnetoelectric (ME) effects, strong magnetostructural correlations,  field-induced spin-flop transitions, and diverse magnetic ground states \cite{Datta, Zheng, Yin, Choi, Narayanan, Maignan, Mehra, Panja, Jana, Goel}. These materials typically crystallize in a centrosymmetric trigonal \textit{P-3c1} structure at ambient conditions. A notable exception is the Ni analogue, Ni$_4$Nb$_2$O$_9$, which stabilizes in  a distorted orthorhombic structure and does not exhibit a linear ME effect in its ferrimagnetically ordered state below $T_C$ = 76 K \cite{Tailleur, Thota}. 

These trigonal A$_4$\textit{B}$_2$O$_9$ compounds exhibit a wide variation in both magnetic ordering ($T_\mathrm{N}$) and Curie–Weiss temperature ($\theta_{P}$).  Upon varying the  $A$-site from Co $\rightarrow$ Fe $\rightarrow$ Mn (with Nb fixed at the $B$-site) leads to a substantial increase in $T_N$ from 27.5 K to 94 K and 120 K, respectively \cite{Jana, Jana2, Jana3, Jana4}. Replacing Nb with Ta, however, moderately suppresses these ordering temperatures to 20.5 K, 85 K, and 110 K, respectively \cite{Maignan, Jana2, Jana3, Jana4}.  A comparable dependence is also reflected in $\theta_{P}$ with Co-based compounds show relatively small values (-51.6 K for Nb and -38.5 K for Ta), whereas Mn-based analogues exhibit much larger negative values (-330 K for Nb and -290 K for Ta), indicating significantly stronger magnetic interactions in the Mn systems \cite{Jana2, Jana3, Jana4}. The microscopic origin of magnetoelectric (ME) coupling varies significantly across this family. Mn-based compounds exhibit collinear spin alignment along the trigonal \textit{c} axis, resulting in a linear ME response primarily governed by magnetostrictive coupling \cite{Zheng, Deng}. In contrast, Co-based systems host an in-plane canted spin structure of Co$^{2+}$ moments, leading to field-induced ferroelectric polarization via the inverse Dzyaloshinskii-Moriya mechanism \cite{Deng2}. Fe-based analogues display spins confined to the \textit{ab} plane with substantially weaker canting, where the ME coupling is mainly driven by magnetostriction and spin--ligand hybridization \cite{Zhang, Ding}.

The nonmagnetic  $B$-site cation further exerts a significant influence on the coupled structural, magnetic, and vibrational properties of the $A_4B_2$O$_9$ family. The substitution of Nb by Ta markedly modifies the structural distortions, phonon dynamics, and magnetoelectric characteristics. For instance, Ueno \textit{et al}. reported a striking contrast in magnetothermal conductivity (MTC) between the Co-based Nb and Ta variants, with the Ta compound showing a large enhancement in MTC compared to the Nb analogue \cite{Ueno}. This pronounced difference was attributed to stronger phonon–magnon interactions mediated by spin–lattice coupling in the Ta compound \cite{Ueno}. Moreover, additional Raman-active modes and anomalous phonon shifts—along with stronger spin–phonon coupling around long-range ($T_{N}$) and short-range magnetic ordering ($T_{sro}$) have been observed in the Ta compound compared to its Nb counterpart, ascribed to the enhanced spin–orbit coupling associated with the heavier Ta cation \cite{Jana2, Jana3, Park}. A similarly strong  $B$-site dependence appears in the Fe-based compounds, where Fe$_4$Ta$_2$O$_9$ (FTO) hosts multiple magnetoelectric and spontaneous type-II multiferroic transitions, in contrast to only two magnetoelectric transitions observed in the Nb analogue \cite{Maignan, Panja}. In our recent low-temperature Raman and magnetic susceptibility study on Mn$_4$Nb$_2$O$_9$ (MNO) and Mn$_4$Ta$_2$O$_9$ (MTO), we identified substantial spin–phonon coupling around both $T_{N}$ and $T_{sro}$ \cite{Jana4}. Notably, contrasting trends in the phonon frequency, linewidth, and integrated intensity of several Raman modes further emphasize the strong influence of the nonmagnetic $B$-site cation on spin–phonon coupling in the Mn-based compounds \cite{Jana4}. 
		
Despite the rich temperature-dependent behavior, only a few high-pressure studies exist on these trigonal ME systems, though the reported results highlight intriguing structural and vibrational phenomena. In Fe$_4$Nb$_2$O$_9$ (FNO), a pressure-induced transition from the trigonal \textit{P-3c1} phase to a cell-doubled monoclinic \textit{C2/c} structure occurs at ~9 GPa, accompanied by signatures of re-entrant ME behavior \cite{Sahu}. This transition closely resembles the low-temperature structural distortion observed near its antiferromagnetic Néel temperature, which coincides with a dielectric anomaly \cite{Jana}. Our earlier high-pressure investigations on Co$_4$Nb$_2$O$_9$ (CNO) and Co$_4$Ta$_2$O$_9$ (CTO) identified three isostructural transitions, followed by distinct sequences of long-range symmetry changes occurring at markedly different pressures in the two systems \cite{Jana2, Jana3}.  Notably, both systems reproduce several low-temperature Raman anomalies—such as the appearance of new Raman modes at comparable frequencies and the renormalization of several modes—indicating a possible correlation between magnetic exchange interactions and lattice vibrations under compression. Both compounds also exhibit pronounced anisotropic lattice compression, with the \textit{c}-axis being more compressible, resulting in a substantial reduction in the \textit{c/a} ratio. A similar pressure-induced anomaly was reported in Mn$_3$NiTa$_2$O$_9$, where unit-cell parameters display irregular behavior near 9.2 GPa, along with a notable decrease in the \textit{c/a} ratio \cite{Singh}. A comparative analysis of the \textit{c/a} ratio and $T_N$ across the \textit{A}$_4$\textit{B}$_2$O$_9$  family reveals a strong correlation between anisotropic lattice modification and magnetic ordering temperature \cite{Jana4}. Specifically, $T_N$ increases as the \textit{c/a} ratio decreases, accomplished by the choice of both magnetic and nonmagnetic cations, highlighting the potential to tune structural and magnetic properties through controlled variation of lattice anisotropy under external perturbation. Thus, pressure-induced structural modification, particularly the anisotropic evolution of the \textit{a} and \textit{c} lattice parameters can serve as an effective means to manipulate magnetic behavior via enhanced spin–lattice coupling \cite{Carlisle, Pawbake}. Since low-temperature Raman studies on MNO and MTO established substantial spin-phonon coupling up to $\sim$250 K, moderate pressures at room temperature are expected to modulate spin–spin interactions through pressure-driven changes in phonon behavior.		
				
Our recent nuclear magnetic resonance (NMR) and Raamn studies on (Mn,Co)$_4$\textit{B}$_2$O$_9$ systems further reveal that Mn-based compounds are more structurally distorted than their Co analogues \cite{Jana4}. Additionally, NMR investigations on Ni$_4$Nb$_2$O$_9$ (NNO) indicate that, despite differences in average crystal symmetry, trigonal MNO closely resembles the orthorhombic NNO at the local scale \cite{Jana5}. This makes Mn-based systems especially compelling for high pressure studies, as they combine pronounced intrinsic local distortions with the highest $T_N$ and $\theta_{P}$ in the \textit{A}$_4$\textit{B}$_2$O$_9$ family. The unit cell further provides a versatile framework for external perturbations, consisting of alternating nearly planar and buckled layers along the \textit{c}-axis (Fig. 1(a)). In the  planar layer, Mn1O$_6$ octahedra form a edge-shared honeycomb network (Fig. 1(b)), while in the buckled layer, alternating edge-shared Mn2O$_6$ and Nb/TaO$_6$ octahedra form two honeycomb sublayers as illustrated in Fig. 1(c)and Fig. 1(e). Face-shared Nb/TaO$_6$ octahedra from adjacent sublayers create Nb$_2$O$_9$ (Ta$_2$O$_9$) units, and corner-shared Mn2O$_6$ octahedra  preserve the buckled framework. The planar and buckled layers are interconnected through face-shared Mn1O$_6$–Mn2O$_6$ and corner-shared Mn1O$_6$–Nb/TaO$_6$ linkages (Fig. 1(d)). The more spatially extended Ta 5\textit{d} orbitals compared to Nb 4$d$ in this structural arrangement are expected to enhance orbital overlaps in MTO, promoting stronger structural distortions and a more pronounced influence on its electronic and magnetic properties under applied pressure \cite{Ueno, Kumar, Somesh}. A comparative high-pressure investigation of the Nb and Ta variants of Mn-based \textit{A}$_4$\textit{B}$_2$O$_9$ is therefore of considerable interest. 
		
\begin{figure*}[ht!]
\includegraphics[trim={1cm 2cm 1cm 1cm},width=16cm]{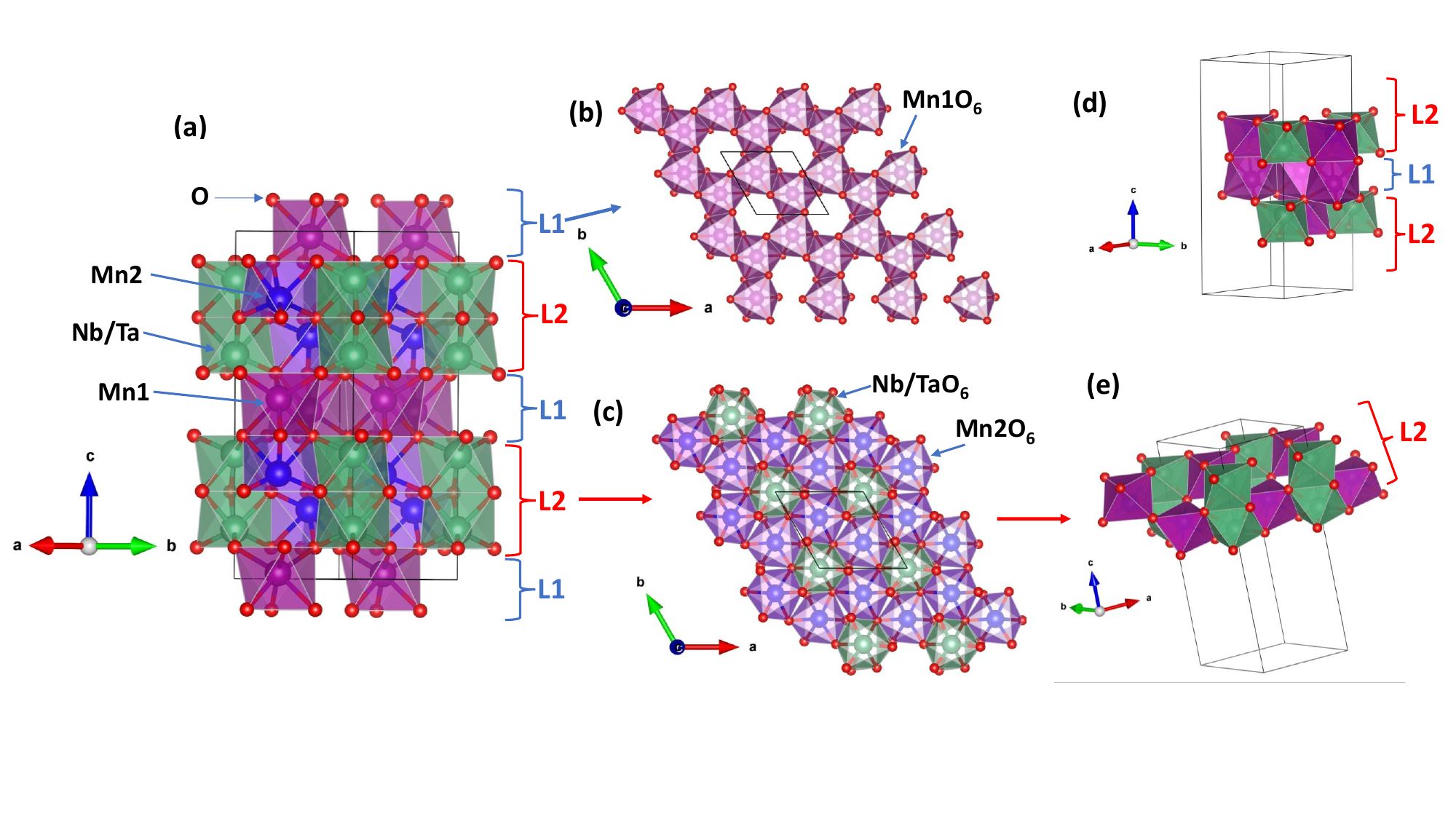}
\caption{\label{fig1} (a) Unit-cell structure of MNO/MTO crystallizing in the trigonal \textit{P-3c1} symmetry at ambient pressure. (b) Planar honeycomb layer (L1) consisting of edge-shared Mn1O$_6$ octahedra, viewed along the $c$ axis. (c) Buckled layer (L2) composed of two honeycomb sublayers formed by edge-shared Mn2O$_6$ and Nb/TaO$_6$ octahedra. (d) Interlayer connectivity between the planar and buckled layers, where Mn1O$_6$ and Mn2O$_6$ octahedra from the respective layers are linked via face sharing, while Nb/TaO$_6$ octahedra from the buckled layer are connected to Mn1O$_6$ octahedra in the planar layer through corner sharing. (e) Three-dimensional view of the buckled layer, illustrating the buckling of corner-shared Mn2O$_6$ octahedra forming two sublayers, with alternating Nb/TaO$_6$ octahedra from adjacent sublayers connected via face sharing.} 
\end{figure*}

In this article, we present a comprehensive study combining high-pressure Raman spectroscopy, and synchrotron x-ray diffraction on MNO and MTO up to $\sim$28 GPa. Multiple isostructural transitions are observed in both systems, manifested through the emergence of new Raman modes and anomalies in the pressure evolution of vibrational and structural parameters, prior to their partial transformation to a monoclinic \textit{P2/c} phase at $\sim$12.5 and $\sim$14 GPa for MNO and MTO, respectively. Both compounds exhibit substantial anisotropic compression, with contraction along the $c$-axis being significantly larger than along the $a$-axis. This effect is much more pronounced than in their Co-based analogues, leading to a rapid reduction in the $c/a$ ratio that drives significant changes in their structural, vibrational, and potentially magnetic properties under high pressure.

\section{Experimental and computational details}

\subsection{High pressure Raman spectroscopy}

High-pressure Raman measurements on polycrystalline samples \cite{Jana4} of MNO and MTO are carried out at room temperature using a Horiba Yvon LabRam HR Evolution spectrometer equipped with a 600 g/mm grating. A 532 nm laser is used for excitation, and the scattered signal is collected in a back-scattering geometry. Static high pressures up to 27.7 GPa are generated using a piston-cylinder-type diamond anvil cell with 400 $\mu$m flat culets. A 200 $\mu$m-thick stainless-steel gasket is pre-indented to 55 $\mu$m, and a 100 $\mu$m central hole is drilled in the pre-indented region using an electric discharge machine to form the sample chamber. Polycrystalline samples, along with a few ruby chips for pressure calibration, are loaded into the chamber, which is subsequently filled with a 4:1 methanol–ethanol mixture as the pressure-transmitting medium. The pressure is determined using the ruby fluorescence method \cite{Mao}.

\subsection{High pressure synchrotron x-ray diffraction}

Synchrotron powder x-ray diffraction (XRD) experiments are carried out on MNO up to 26.5 GPa and MTO up to 9 GPa at the Xpress beamline of Elettra Synchrotron (Trieste, Italy). A monochromatic x-ray beam with a wavelength of 0.5 \AA~ and 50~$\mu$m width is employed \cite{Lotti}. Pressures are generated using a symmetric membrane driven diamond anvil cell  with 400 $\mu$m flat culets. Fine powder samples, along with ruby balls as pressure markers, are loaded into a 150 $\mu$m-diameter sample cavity prepared on a pre-indented stainless-steel gasket. A 4:1 methanol–ethanol mixture is used as the pressure-transmitting medium to maintain hydrostatic condition. Two-dimensional  (2D) diffraction images are collected on a large-area PILATUS3 S 6M detector placed at a distance 92.6 cm from the sample. Calibration of the sample-to-detector distance and beam geometry is performed using LaB$_6$ patterns in Dioptas software \cite{Prescher}.  Integration of the 2D rings is also carried out using Dioptas. Preliminary lattice parameter estimation and space group indexing are performed using CRYSFIRE/DICVOL and CHECKCELL \cite{Shirley, Boultif, Laugier}, followed by detailed structural refinement using Le Bail or Rietveld methods with EXPGUI–GSAS and FullProf software packages \cite{Toby, Carvajal}. 

An additional high-pressure run for MTO up to 25 GPa is conducted at the BL-12 beamline of Indus-2 Synchrotron, RRCAT, Indore. For this experiment, polycrystalline MTO along with silver powder (as a pressure calibrant) and 4:1 methanol–ethanol pressure medium are loaded into a 100 $\mu$m sample cavity of a DAC with 300 $\mu$m diamond culets. A monochromatic X-ray beam of 0.7 \AA~ wavelength and 80 $\mu$m diameter is used, and diffraction patterns are recorded on a MAR 3450 image plate detector. Pressures inside the sample cavity are determined from the P–V equation of state of silver \cite{Dewaele}.

\subsection{Density functional theory computation} 

First-principles calculations were performed within the framework of density functional theory (DFT) using the Vienna Ab initio Simulation Package to investigate the structural behaviour of the system under applied pressure \cite{Kresse, Kresse2, Kresse3}. The exchange–correlation effects were treated using the Generalised Gradient Approximation in the form Perdew–Burke–Ernzerhof functional \cite{Perdew}. The electronic wavefunctions were expanded using a plane-wave basis set with a kinetic energy cutoff of 700 eV. Structural optimisation was performed until the total energy converged within 10$^{-7}$ eV, while the residual Hellmann–Feynman forces acting on each atom were reduced below 0.001 eV\AA$^{-1}$. Brillouin zone integrations were carried out using a Gamma scheme with a 7 $\times$ 7 $\times$ 2 k-point mesh for the ambient \textit{P-3c1} phase \cite{Monkhorst}.

\section{Results and Discussion}

\subsection{High pressure Raman studies on MNO}

Ambient Raman spectra of MNO and MTO are shown in Fig. S1 of the Supplementary Material (SM) \cite{SM}. Although, nearly equal (17 for MNO and 16 for MTO) Raman-active modes are observed in both systems, they exhibit distinct behavior in terms of mode positions, lineshapes, and intensities, reflecting the complex mechanisms discussed elsewhere \cite{Jana4}. Figure 2 presents high-pressure Raman spectra of MNO up to 15.5 GPa at selected pressures in the 70–700 cm$^{-1}$ range. As expected for a compressed unit cell, all Raman modes exhibit a blue shift with increasing pressure. Notably, around 2 GPa, a weak Raman mode appears between the E$_g$(4) and A$_{1g}$(1) modes at ~199 cm$^{-1}$, labeled as $\Omega$(1), suggesting local symmetry breaking (Fig. 2). At the same pressure, the E$_g$(8) mode, previously merged with E$_g$(7), begins to separate, indicating increased structural distortion. A similar behavior is observed at 248 K under ambient pressure, close to the emergence of short-range magnetic correlations \cite{Jana4}. The E$_g$(2) mode, which has comparable intensity to the neighboring E$_g$(3) mode at ambient pressure, becomes less intense at 2 GPa and gradually decreases relative to E$_g$(3) at higher pressures. This evolution is in contrast to its Co analogue, where comparable behavior is observed at a significantly higher pressure of 11.5 GPa \cite{Jana2}. Interestingly, the $\Omega$(1) mode also appears at 11.5 GPa in CNO \cite{Jana2}, suggesting that pressure-induced structural and vibrational modifications occur at much lower pressures in MNO, likely due to its more distorted local structure \cite{Jana4}. Notably, a similar low-frequency mode splitting is reported in NNO at 2.1 GPa, which shares a comparable local structure with MNO as revealed by NMR studies \cite{Jana4, Jana5}.

\begin{figure*}[ht!]
\includegraphics[trim={1cm 4cm 2cm 1cm},width=18cm]{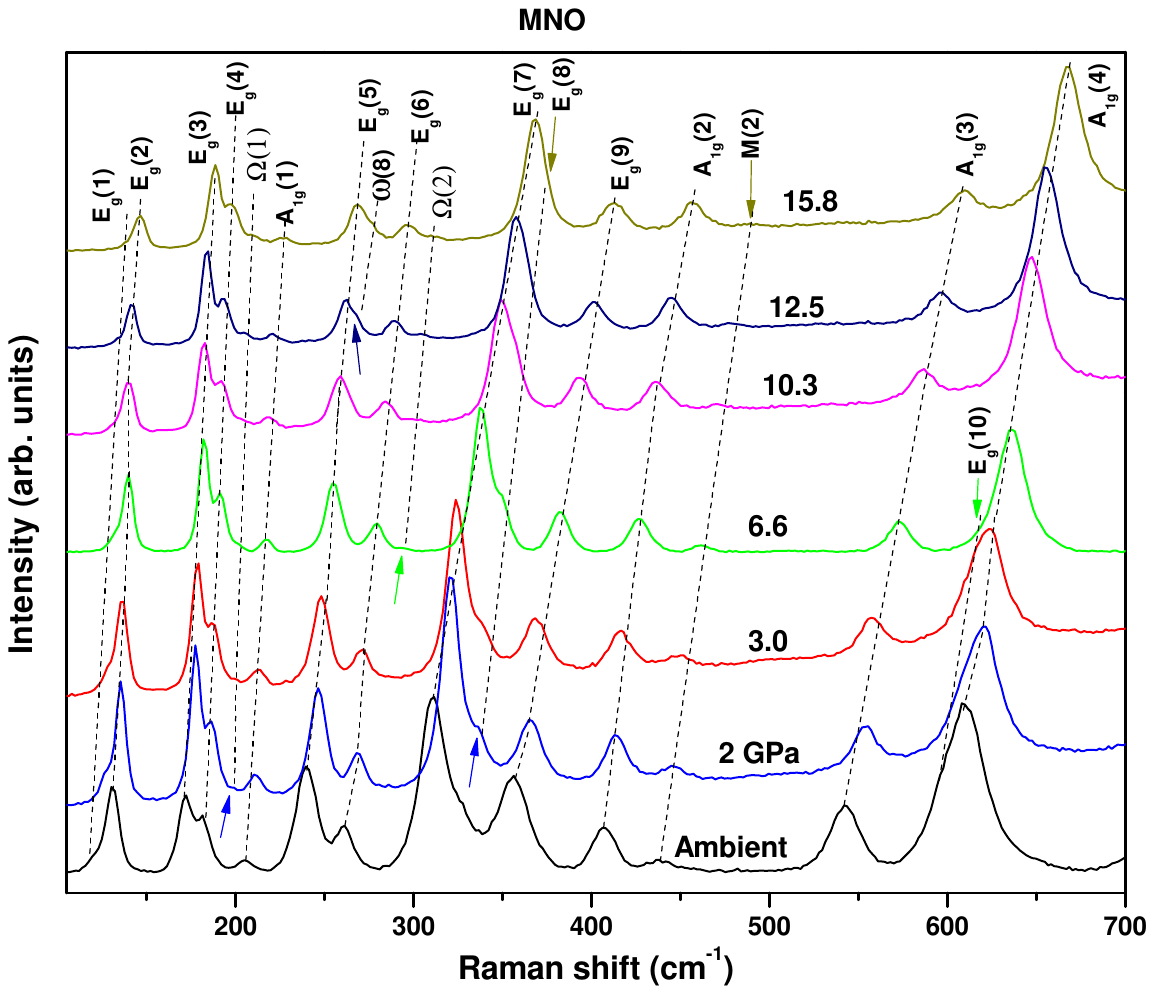}
\caption{\label{fig2} High-pressure Raman spectra of MNO up to 15 GPa in the wavenumber range 105–700 cm$^{-1}$. The emergence of new Raman modes is indicated by upward arrows, while the disappearance of modes is marked by downward arrows. The pressure evolution of individual modes is tracked by dashed black lines. } 
\end{figure*}

The E$_g$(10) mode initially shows separation trend from the overlapping A$_{1g}$(4) mode up to 2 GPa, mirroring its low-temperature behavior. However, above 2 GPa, it exhibits the opposite trend, fully merging with A$_{1g}$(4) by 6.6 GPa. At this pressure, an additional mode, $\Omega$(2), emerges at around 294 cm$^{-1}$, as highlighted by the green arrow in Fig. 2. Interestingly, this mode also appears at 77 K, well below the long-range magnetic ordering temperature, in our low-temperature Raman study \cite{Jana4}, indicating potential structural and magnetic correlations. Beyond 6.6 GPa, the E$_g$(8) mode gradually converges with E$_g$(7). At 12.5 GPa, the E$_g$(5) mode splits into two distinct features: E$_g$(5) and $\omega$(8). By 15.5 GPa, E$_g$(8) is fully merged into E$_g$(7), and the M(2) mode is nearly suppressed.

\begin{figure*}[ht!]
\includegraphics[trim={1cm 4cm 2cm 1cm},width=18cm]{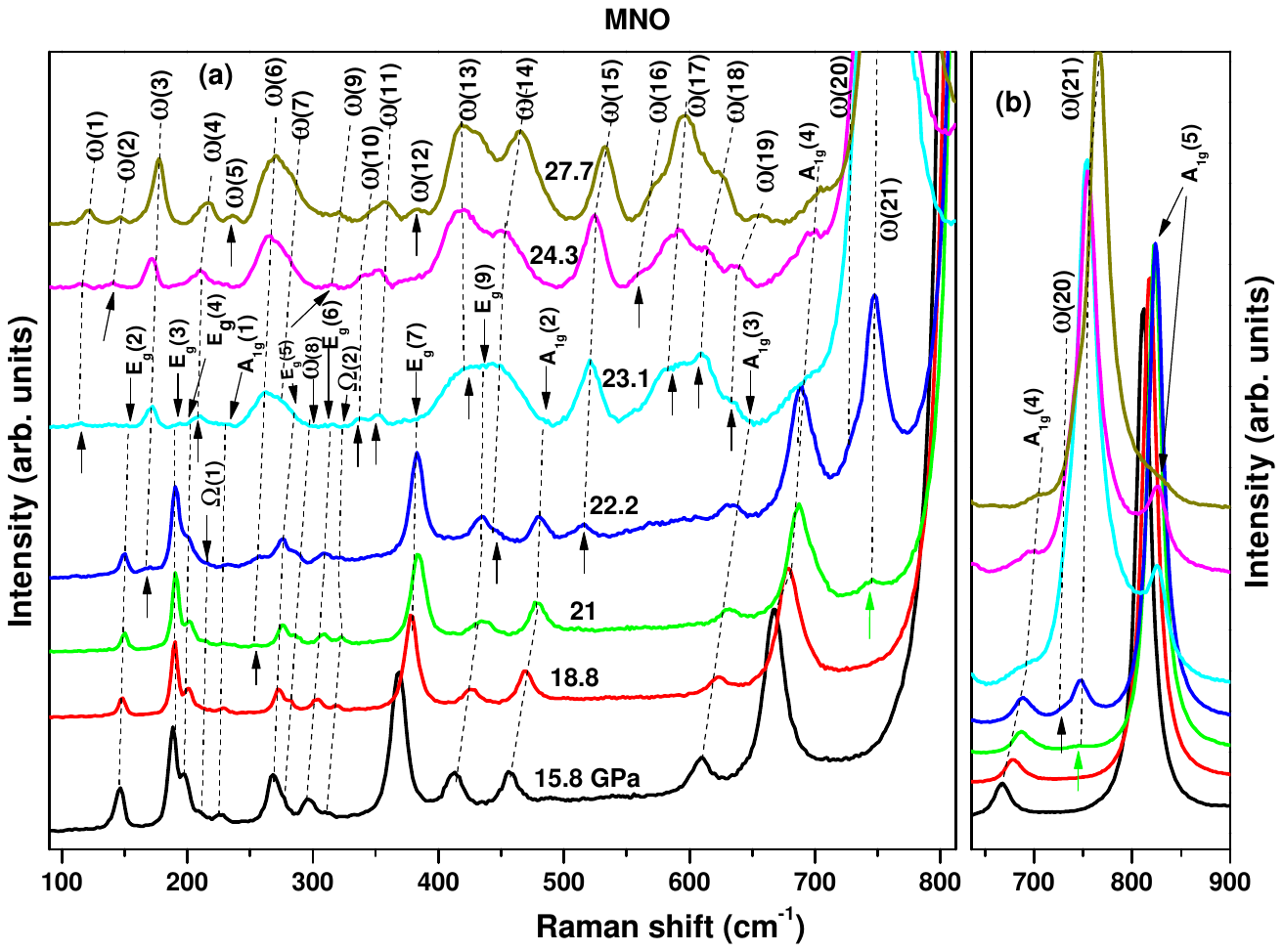}
\caption{\label{fig3} High-pressure Raman spectra of MNO over the pressure range 15–27 GPa in the wavenumber ranges (a) 90–810 cm$^{-1}$ and (b) 630–840 cm$^{-1}$. The appearance and disappearance of Raman modes are indicated by upward and downward arrows,  respectively. The pressure evolution of individual modes is traced by black dashed lines.} 
\end{figure*}

Figure 3 presents the high-pressure Raman spectra of MNO in the range 15.8–27.7 GPa over 90–900 cm$^{-1}$. Above 15.8 GPa, the intensity of the E$_g$(2) mode is significantly reduced, while the remaining modes show only minor changes up to 21 GPa. At this pressure, an octahedral mode emerges around 743 cm$^{-1}$ ($\omega$(21)), as indicated by the green arrow in Fig. 3(a)-(b). The activation of such an octahedral mode between A$_{1g}$(4) and A$_{1g}$(5) has been associated with a long-range structural transition in CNO and CTO, as reported in Refs. \cite{Jana2, Jana3}. Additionally, a new low-frequency mode appears at 21 GPa around 254 cm$^{-1}$, designated as $\omega$(6). With a slight pressure increase to 22.2 GPa, the $\Omega$(1) mode, which was activated at 2 GPa, disappears, while three new modes emerge at 168 ($\omega(3)$), 444 ($\omega(14)$) and 517 ($\omega(15)$) cm$^{-1}$. Simultaneously, the intensity of  $\omega$(21) mode increase substantially, and by 23.1 GPa, this mode surpass the neighboring A$_{1g}$(5), the strongest mode of the ambient phase (Fig. 3(b)). At this pressure, a pronounced renormalization of the Raman spectrum is observed. Several modes of the ambient-pressure \textit{P-3c1} phase, including E$_g$(2), E$_g$(3), E$_g$(5), A$_{1g}$(1), E$_g$(6), E$_g$(7), E$_g$(9), and A$_{1g}$(3), are completely suppressed, as highlighted by downward arrows in Fig. 3(a), while multiple new modes appear, marked by upward arrows, signaling a major structural transformation. Interestingly, similar dramatic modifications, appearance and disappearance of multiple modes along with the newly emerged octahedral mode surpassing the strongest ambient-phase mode are observed in NNO at the same pressure \cite{Jana5}. In contrast, in CNO, although the octahedral region shows intensity changes, no such emergence or disappearance of multiple modes is reported \cite{Jana3}. This suggests that the similar local structures of MNO and NNO likely play a dominant role in the observed pressure-induced structural and vibrational transformations.

	On further increase of pressure additional Raman modes appear at 24.3 and 27.7 GPa, for instance, $\omega$(2), $\omega$(9), and $\omega$(16) at 24.4 GPa, and $\omega$(5) and $\omega$(12) at 27.7 GPa. In the pressure range of 23.1–27.7 GPa, all the newly activated Raman modes exhibit a continuous increase in intensity, becoming sharp and well-defined. This behaviour contrasts with that of the Co analogue, where the Raman modes associated with the high-pressure phase remain relatively weak. Well-resolved and comparably intense Raman modes corresponding to the high-pressure phase are also reported in NNO \cite{Jana5} at similar pressure ranges, indicating that the high-pressure behaviour of MNO closely resembles that of NNO. Notably, the strongest and second-strongest Raman modes of the ambient phase, A$_{1g}$(4) and A$_{1g}$(5), persist weakly up to 27.7 GPa, suggesting the coexistence of a minor fraction of the ambient \textit{P-3c1} phase.

	To gain deeper insight into the lattice dynamics, all Raman spectra of MNO are fitted using Lorentzian line shapes. The extracted frequencies and linewidths of the modes up to 315 cm$^{-1}$ are presented in Fig. 4. The three low-frequency Raman modes, E$_g$(2)-E$_g$(4), are particularly sensitive to applied pressure. These modes exhibit distinct slope changes in their frequency evolution at approximately 2, 6.6, and 10 GPa (Figs. 4(a)-(c)), indicative of possible isostructural transitions or local structural rearrangements. They  gradually hardens up to 2 GPa becomes softer at 2 GPa relative to their lower pressure variations and start to soften slightly above 6.6 GPa. In contrast, above 10 GPa, they start to harden again. The linewidths of these modes gradually narrow up to 2 GPa, then exhibit slow decrease and produce a shallow minimum around 6.6 GPa, followed by broadening up to 10 GPa. Beyond 10 GPa, their linewidths decrease up to about 12.5 GPa. 
	
	\begin{figure*}[ht!]
\includegraphics[trim={1cm 6cm 3cm 1cm},width=18cm]{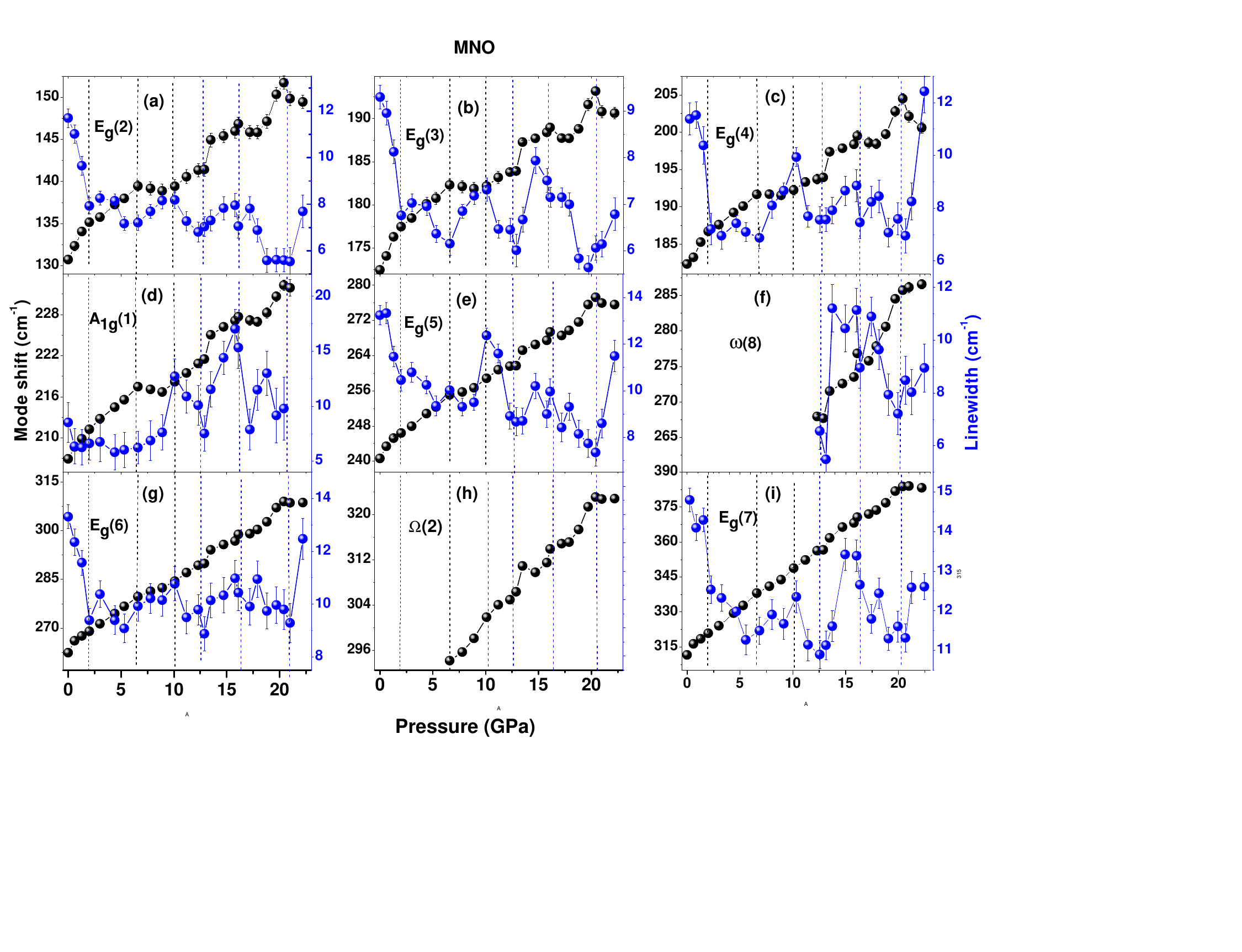}
\caption{\label{fig4} (a)-(i) Pressure evolution of the Raman mode frequencies and linewidths of low-frequency modes in MNO. Dashed black lines indicate the isostructural transition pressures, while blue dashed lines mark the onset of long-range structural transitions or major structural rearrangements. } 
\end{figure*}

	A marked frequency jump of the E$_g$(2)–E$_g$(4) modes occurs near 12.5 GPa, accompanied by a minimum in their linewidths, signaling the onset of a long-range structural transformation (Figs. 4(a)-(c)). This behaviour closely parallels the reported frequency discontinuity and linewidth anamoly of several Raman modes  around 13 GPa in NNO, associated with a \textit{Pbcn} to partially \textit{P2/c} transition \cite{Jana5}. Furthermore, additional irregularities in both frequency and linewidth are observed around 15.8 and 21 GPa, which likely reflect changes in structural rearrangements or symmetry transitions under increasing pressure.

	For the A$_{1g}$(1) and E$_g$(5) modes, a minor slope change in frequency and slight irregularity in their linewidth are observed around 2 GPa, as shown in Figs. 4(d)–(e). In the pressure range of 6.6–10 GPa, the A$_{1g}$(1) mode exhibits a red-shift trend accompanied by a marked broadening of its linewidth. While the linewidth of the E$_g$(5) mode develops a peak-like feature near 10 GPa (Fig. 4(e)). Beyond 10 GPa, the A$_{1g}$(1) mode reverts to a blue-shift behavior up to 12.5 GPa, where it shows a distinct frequency jump and a minimum in linewidth, mirroring the behavior of the low-frequency E$_g$(2)–E$_g$(4) modes. A moderate frequency jump with a concurrent linewidth minimum is also observed for the E$_g$(5) mode at the same pressure. At 15.8 GPa, the A$_{1g}$(1) mode shows a sharp slope change in frequency and linewidth maximum before weakening significantly. The E$_g$(5) mode shows a minor frequency discontinuity at the same pressure, followed by a clear slope change and rapid linewidth broadening at 21 GPa. The newly activated $\omega$(8) mode, which appears at 12.5 GPa, displays pronounced slope changes in frequency evolution and strong anomalies in its full width at half maximum (FWHM) around 15.8 and 21 GPa, suggesting strong structural modifications under high pressure.

	The E$_g$(6) mode exhibits a noticeable discontinuity in frequency around 12.5 GPa and a clear slope change around 21 GPa (Fig. 4(g)). In contrast, the E$_g$(7) mode shows a monotonic increase in frequency with pressure up to 21 GPa, where it displays a pronounced slope change. The linewidth evolution of both modes reveals distinct anomalies at approximately 2, 6.6, 10, 12.5, and 15.8 GPa, indicating pressure-induced modifications in lattice dynamics and possible correlation with magnetic interactions. The $\Omega$(2) mode, which emerges at 6.6 GPa, exhibits a pronounced frequency jump at 12.5 GPa, followed by another slope change in its frequency evolutio around 21 GPa, further supporting the occurrence of multiple structural or dynamical transformations under compression. Its linewidth is not quantitatively extracted owing to its weak intensity and poorly defined peak profile.  

	The pressure evolution of higher-frequency Raman modes is presented in Fig. S2 \cite{SM}. A finite softening in frequency and a discontinuity in the gradual linewidth narrowing can be observed in the E$_g$(8) mode above 2 GPa, compared to its 0–2 GPa trend (Fig. S2(a)). An additional minimum in the linewidth profile of this mode is seen, concurrent with a slight slope change in its frequency around 6.6 GPa. The frequencies of E$_g$(9), and A$_{1g}$(2)-A$_{1g}$(5) do not show significant peculiarities as shown in Figs. S2(b)-(f). They only exhibit slight anomalies around 21 GPa. Additionally, A$_{1g}$(2) and A$_{1g}$(5) undergo moderate discontinuities around 12.5 GPa during the long-range structural transitions. 
	
	This behavior suggests that the mode shift of the high-frequency Raman modes are primarily associated with typical unit cell compression. However, notable peculiarities are observed in their linewidth variations with pressure. In particular, the linewidth of the A$_{1g}$(2) mode gradually narrows up to 2 GPa, then slightly broadens between 2 and 6.6 GPa, before rapidly decreasing up to 10 GPa (S2(c)). The linewidth of the A$_{1g}$(5) mode sharpens up to 2 GPa, followed by a much slower rate of narrowing between 2 and 6.6 GPa. At 6.6 GPa, it experiences an upward jump before decreasing at a much slower rate, remaining almost unchanged above 12.5 GPa. The linewidth of the E$_g$(9) and A$_{1g}$(4) modes decreases up to 10 GPa, with a moderate change in rate at 2 GPa, and then both modes start to broaden. The A$_{1g}$(3) mode also exhibits strong peculiarities in its linewidth behavior around 6.6 GPa, though no noticeable change in its frequency is observed. These behaviors of high-frequency Raman modes, with stronger peculiarities in linewidth than in frequency, suggest that other mechanisms, such as changes in electronic and magnetic properties, are involved in the lattice vibrations under pressure.

\subsection{High pressure Raman study on MTO}
	
	High-pressure Raman spectra of MTO up to 14.6 GPa  are presented in Fig. 5 in the range of 70–775 cm$^{-1}$. Upon a slight increase in pressure to 0.5 GPa, a remarkable pressure effect is evident in the Raman spectrum. The Raman modes above 200 cm$^{-1}$ exhibit a typical blue shift, whereas those below 200 cm$^{-1}$ display an unusual red shift. At this pressure, two additional modes, $\nu$(1) and $\nu$(2), emerge around 198 and 225 cm$^{-1}$, respectively, suggesting local symmetry breaking. Concurrently, the E$_g$(6) mode, previously merged with E$_g$(5), becomes clearly resolved, and the E$_g$(9) mode begins to separate from the neighboring E$_g$(8) mode. Notably, the emergence of the $\nu$(1) and $\nu$(2) modes has also been reported at 248 K and 108 K, corresponding to short-range spin correlations and long-range magnetic ordering, respectively \cite{Jana4}. Interestingly, the activation of the $\nu$(1) mode is further observed in the $A$-site variant MNO at 2 GPa in the present study and in the $B$-site variant CTO at 3 GPa as reported in Ref. \cite{Jana3}. The close analogy between the pressure-induced effects observed at 0.5 GPa in MTO, the low-temperature behavior of the same system, and the high-pressure responses of MNO and CTO suggests a robust coupling between structural and magnetic degrees of freedom in MTO, reflecting its strong sensitivity to external pressure.

	\begin{figure*}[ht!]
\includegraphics[trim={1cm 4cm 3cm 0cm},width=16cm]{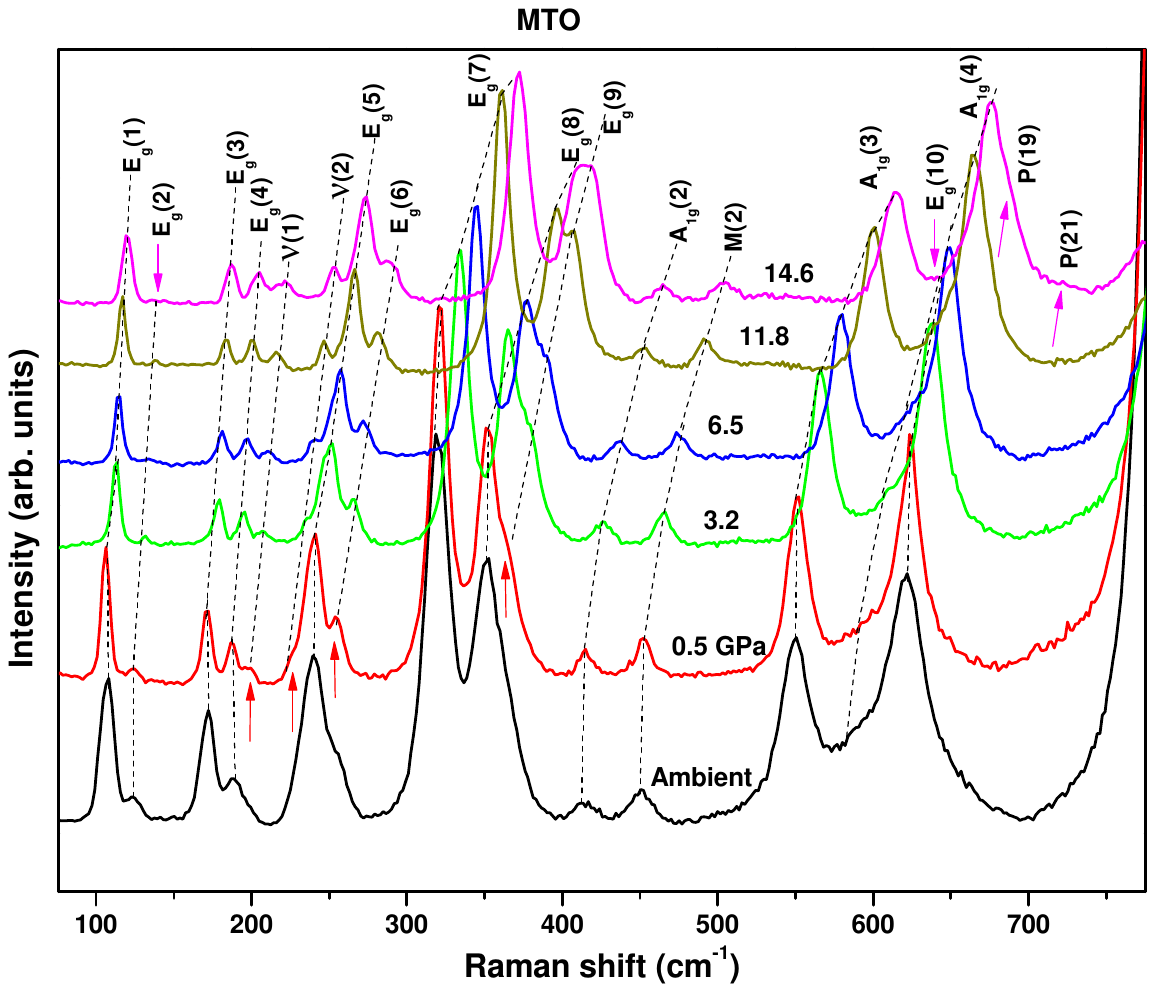}
\caption{\label{fig1} High-pressure Raman spectra of MTO up to 15 GPa over the frequency range 75–775 cm$^{-1}$. The appearance/separation and disappearance of Raman modes are indicated by upward and downward arrows, respectively. Black dashed lines trace the pressure evolution of the individual modes. } 
\end{figure*}

	Upon further compression, the $\nu$(2) mode becomes well defined at 3.2 GPa, with its intensity increasing gradually with pressure. In contrast, the intensity of the $\nu$(1) mode initially decreases slightly around 3.2 GPa and then increases steadily under higher compression. The E$_g$(2) and E$_g$(10) modes progressively lose intensity and completely vanish near 14.6 GPa, suggesting substantial pressure-induced structural modifications. Consistent with this evolution, two high-frequency octahedral modes, labeled P(19) and P(21), emerge at 14.6 GPa around 688 and 724 cm$^{-1}$, respectively (Fig. 5). Their appearance signals the onset of a new high-pressure phase coexisting with the ambient trigonal \textit{P-3c1} structure. Notably, low-temperature studies \cite{Jana4} also reveal the \(P(19)\) mode near the long-range magnetic ordering temperature, hinting at a strong magnetostructural coupling. The pressure evolution of neighboring E$_g$(7), E$_g$(8), and E$_g$(9) modes exhibit particularly intriguing behavior. It is noteworthy that, among these modes, the latter two display intensity transfer at ambient conditions between the Nb- and Ta-based analogues (MNO and MTO), as shown in Fig. S1 and discussed in our recent work \cite{Jana4}. The E$_g$(9) mode, which begins to separate from E$_g$(8) with higher initial intensity, continues to gain relative strength upon compression. At approximately 11.8 GPa, these two modes become well resolved, and at higher pressures their intensities become nearly comparable. Eventually, they merge into a single broad feature with a flat-top profile at 14.6 GPa, signifying the coexistence of two modes of equal intensity. Throughout this evolution, the Eg(7) mode remains spectrally isolated and does not interact with the other two modes. This behavior contrasts sharply with that of MNO, where the E$_g$(8) mode starts to separate from E$_g$(7) above 2 GPa but retains significantly lower intensity, ultimately merging with E$_g$(7) around 15.8 GPa, while the E$_g$(9) mode remains largely unaffected by this transformation process. 

	High-pressure Raman spectra of MTO up to 27.4 GPa are shown in Fig. 6. Above 14.6 GPa, several pronounced spectral modifications emerge, indicating substantial rearrangement within the unit cell structure. The intensity of the E$_g$(8) mode gradually decreases relative to the neighboring E$_g$(9) mode and eventually vanishes above 21.2 GPa. At this pressure, eight additional Raman modes become activated, as marked by green arrows in Figs. 6(a) and 6(b), evidencing strong structural reorganization in MTO. Concurrently, the intensity of the octahedral mode P(21) increases sharply starting at 21.2 GPa. In addition, a hump-like excitation (E), spanning the 80–110 cm$^{-1}$ range, begins to develop around this pressure (green dashed arrow in Fig. 6(a)). Interestingly, a similar feature has been observed near the long-range magnetic ordering temperature as reported in our recent low-temperature study\cite{Jana4}, suggesting that this excitation may have a magnetic origin.

\begin{figure*}[ht!]
\includegraphics[trim={1cm 4cm 2cm 1cm},width=18cm]{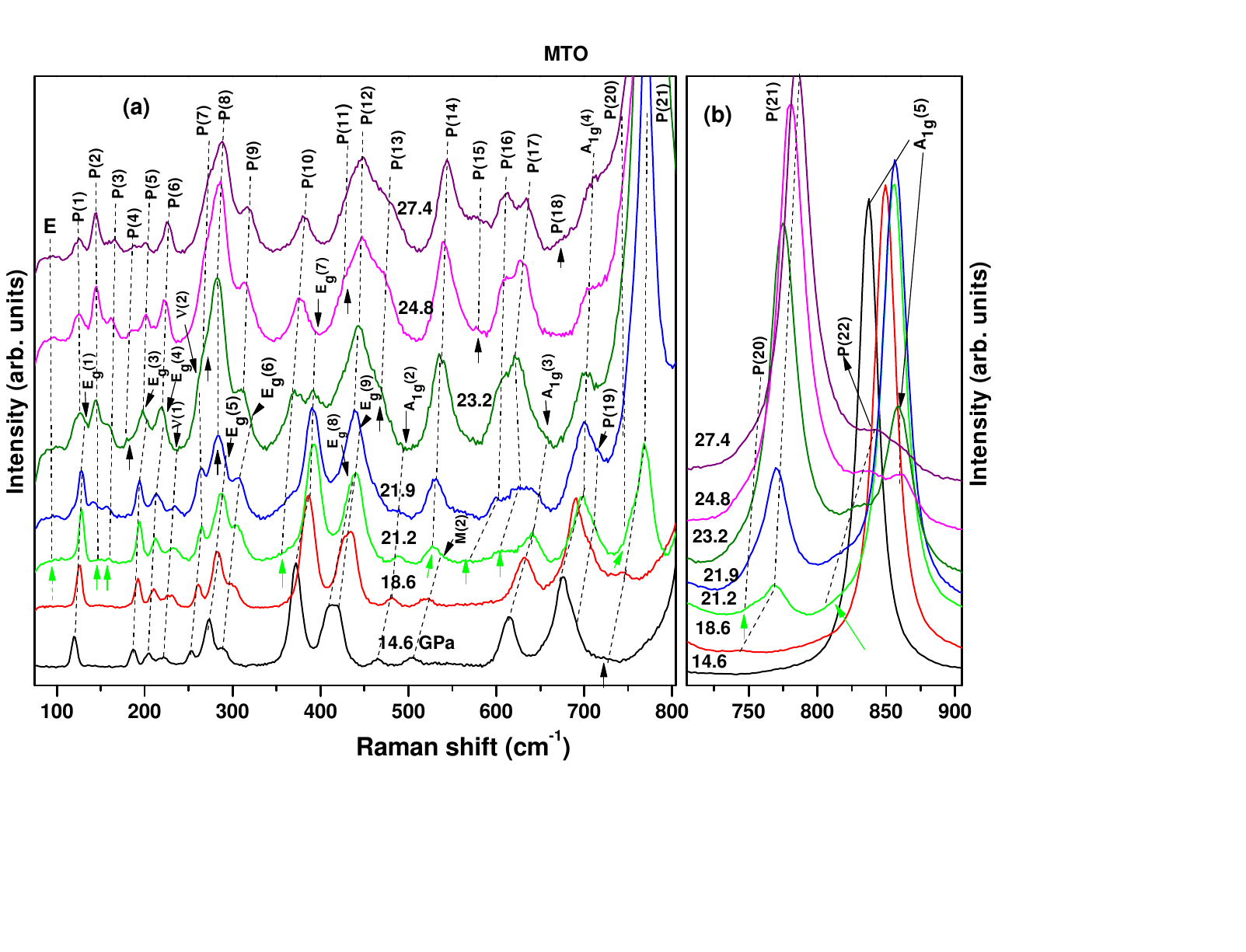}
\caption{\label{fig5} High-pressure Raman spectra of MTO over the pressure range 15–27 GPa in the frequency ranges (a) 75–805 cm$^{-1}$ and (b) 705–905 cm$^{-1}$. The appearance and disappearance of Raman modes are indicated by upward and downward arrows, respectively. The pressure-induced evolution of individual modes is traced by black dashed lines.} 
\end{figure*}

	Above 21.2 GPa, the newly emerged Raman modes grow rapidly in intensity and become prominent by 23.2 GPa, while several modes characteristic of the ambient \textit{P-3c1} phase are either suppressed (downward arrows) or evolve into distinct modes associated with the high-pressure phase. With the appearance of three additional modes, the Raman spectrum at 23.2 GPa becomes entirely different from that below 21 GPa, except for the persistence of the A$_{1g}$(4) and  A$_{1g}$(5) modes, signifying that a major fraction of the material has transformed into the high-pressure phase. At 24.8 GPa, the coexistence of mode emergence and disappearance (upward and downward arrows in Fig. 6(a)) further confirms ongoing structural evolution. Between 21.9 and 24.8 GPa, the P(21) mode intensity increases rapidly, while the strongest ambient-phase mode, A$_{1g}$(5), weakens substantially (Fig. 6(b)). Consequently, the P(21) mode becomes the most intense Raman feature by 23.2 GPa, indicating the dominance of the high-pressure phase over the ambient \textit{P-3c1} phase. At 27.4 GPa, the A$_{1g}$(4) and A$_{1g}$(5) mode are barely detectable, reflecting the minor presence of the ambient \textit{P-3c1} phase.

	The pressure evolution of the mode frequency and linewidth for Raman modes of MTO below 360 cm$^{-1}$  is presented in Fig. 7. Among all the observed modes, the three low-frequency modes E$_g$(1), E$_g$(3), and E$_g$(4) are most strongly influenced by pressure. Upon increasing pressure from ambient to 0.5 GPa, these modes exhibit clear softening, as shown in Fig. 7(a)-(c). Above 0.8 GPa, they undergo a rapid blue shift up to 3.2 GPa. With further compression, the frequencies of these modes continue to increase gradually up to 13 GPa, accompanied by a slight change in slope near 10 GPa. Around 14 GPa, a pronounced jump in the mode frequencies of all three modes is observed, similar to the behavior of MNO at 12.5 GPa. Notably, the appearance of two octahedral modes near this pressure coincides with the frequency discontinuity, indicating the onset of a long-range structural transition. In the 14.6–21 GPa range, these modes exhibit a more pronounced blue shift characterized by a distinctly steeper slope compared to the 3.2–13 GPa regime. At around 18.6 GPa, the E$_g$(3) mode frequency exhibits a pronounced irregularity, coinciding with the clear visibility of the octahedral mode P(21), which starts to emerge at 14.6 GPa.

\begin{figure*}[ht!]
\includegraphics[trim={0cm 5cm 0cm 1cm},width=20cm]{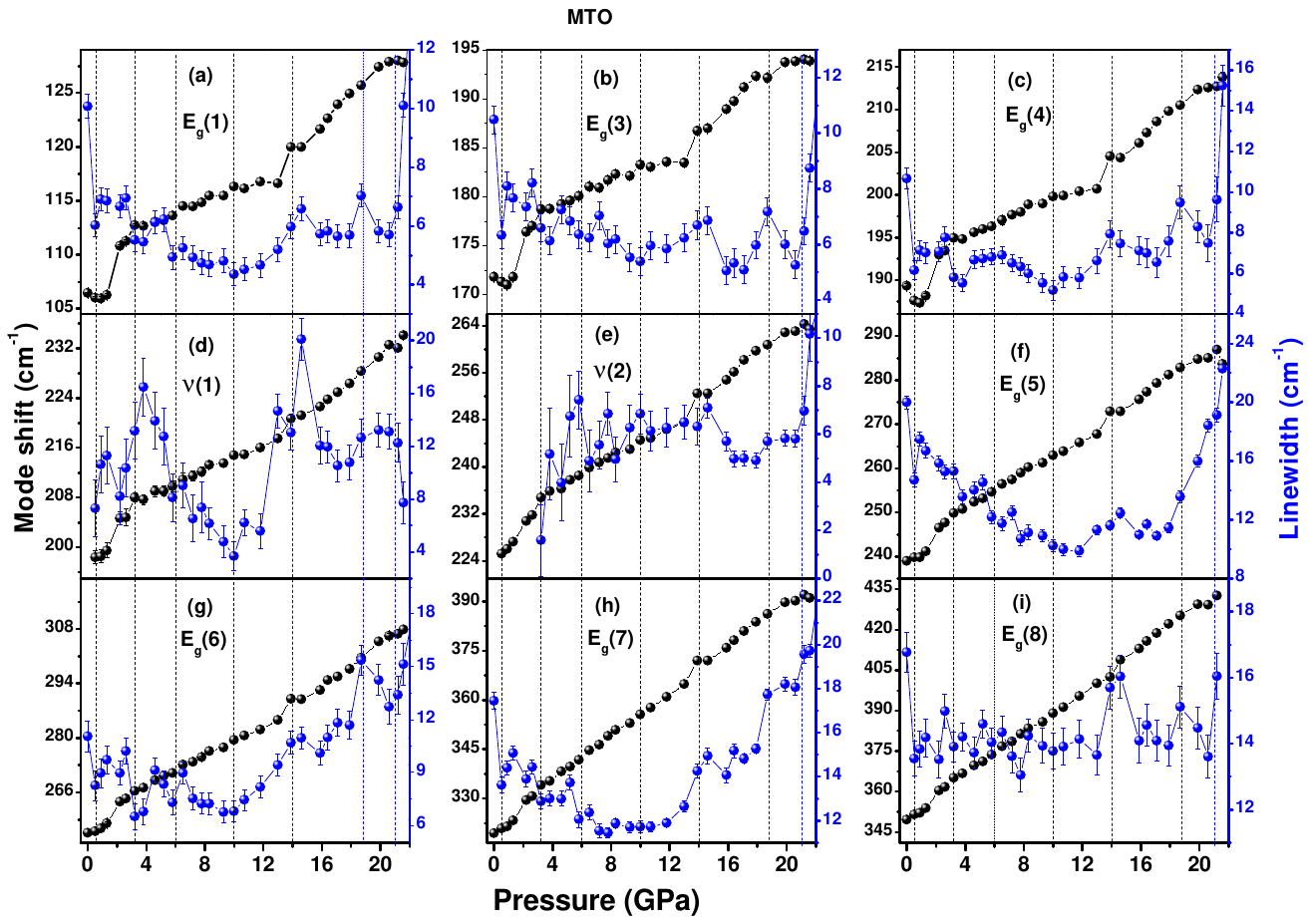}
\caption{\label{fig1} (a)-(i) Evolution of the low-frequency Raman modes (up to 360 cm$^{-1}$) of MTO as a function of pressure. The vertical black dashed lines highlight isostructural transitions, whereas blue dashed lines mark the beginning of long-range structural transition or major structural transformations or ordering.} 
\end{figure*}

Anomalous behavior in the linewidths of the three Raman modes E$_g$(1), E$_g$(3), and E$_g$(4) is also observed at the same pressure values where slope changes in mode frequencies occur. The linewidths of these modes initially narrow as pressure increases from ambient to 0.5 GPa, followed by a slight broadening at 0.8 GPa (Fig. 7(a)-(c)). Between 0.8 and 2.6 GPa, the linewidths remain nearly constant within experimental uncertainty, then decrease noticeably at 3.2 GPa. Additional anomalies are evident around 6, 10 GPa, 14 GPa, and 18.6 GPa. Above 21 GPa, the linewidths exhibit a rapid broadening trend, while the mode frequencies of E$_g$(1) and E$_g$(3) display a softening tendency, suggesting that these modes begin to evolve into distinct vibrational features of the high-pressure phase. This behavior is consistent with the pronounced spectral transformations observed in the 21.9–23.2 GPa range. The slope changes in mode frequency and the anomalous behavior in linewidth observed at 0.5, 3.2, 6, and 10 GPa, highlighted by black dashed lines, are likely associated with isostructural transitions,  potentially signaling underlying changes in electronic and magnetic properties. In contrast, the anomalies appearing around 14, 18.6, and 21 GPa, marked by blue dashed lines, signify long-range structural transitions or rearrangements of the unit cell driven by the competition between the high-pressure and ambient phases. Similar trends have been observed in MNO as well as in the Co-based analogues CNO and CTO \cite{Jana2, Jana3}.

The $\nu$(1) mode, which emerges at 0.5 GPa, exhibits distinct slope changes in its frequency evolution around 3.2 GPa and 14 GPa before disappearing above 21 GPa (Fig. 7(d)). Its linewidth also shows pronounced anomalies, displaying two maxima around 3.2 GPa and 14 GPa, and a minimum around 10 GPa. An additional slight irregularity is observed at around 6 GPa in the linewidth behaviour of the $\nu$(1) mode.  Similar to the lower-frequency modes, the $\nu$(2) mode shows a pronounced hardening up to 3.2 GPa, beyond this pressure, the hardening rate decreases significantly, followed by an abrupt frequency jump at 14 GPa (Fig. 7(e)). Its linewidth evolution reveals clear irregularities around 6, 14 GPa and 21 GPa. A distinct discontinuity in both frequency and linewidth is also observed in the E$_g$(5)–E$_g$(7) modes near 14 GPa, corresponding to the onset of the long-range structural transition as depicted in Figs. 7(f)-(h). In addition, the linewidth of the E$_g$(6) mode narrows significantly around 3.2 GPa, then slightly broadens with increasing pressure and narrows again above 6 GPa exhibiting a minimum near 10 GPa. Beyond this pressure its linewidth broadens progressively making three irregularities at around 14,  18.6, and 21 GPa. The frequency of the E$_g$(8) mode shows an almost monotonic hardening with a subtle discontinuity around 14 GPa; however, its FWHM displays distinct anomalies, a sharp decrease at 0.5 GPa and a pronounced peak at 14 GPa, indicating the influence of the first isostructural and subsequent long-range structural transition.

	The pressure evolution of the higher-frequency Raman modes of MTO is presented in Fig. S3 \cite{SM}. The mode frequencies of E$_g$(9), M(2), and A$_{1g}$(2)–A$_{1g}$(5) exhibit moderate discontinuities around 14 GPa, whereas the E$_g$(10) mode vanishes and the P(21) mode emerges at this pressure. In terms of linewidth evolution, A$_{1g}$(3)–A$_{1g}$(5) modes show a pronounced drop at 0.5 GPa, while the E$_g$(10) mode reaches a minimum around this pressure. At higher pressures the linewidth of A$_{1g}$(3) remains nearly constant up to 14 GPa, beyond which it begins to broaden significantly. Above 3.2 GPa, the linewidth of the E$_g$(10) mode gradually decreases, with the rate of narrowing becoming more pronounced beyond 10 GPa. In contrast, the A$_{1g}$(5) mode exhibits a steady broadening above 10 GPa up to 23 GPa, accompanied by two distinct irregularities around 14 GPa and 21 GPa. The octahedral mode P(21) emerged at the onset of strcutural transition around 14 GPa, exhibits slope change in its mode frequency and linewdth peculiarity around 18.6 and 21 GPa, while the other octahedral mode P(22), activated at 21 GPa, progressively hardens and broadens with pressure.

\subsection{High pressure synchrotron XRD studies on MNO}

	High-pressure powder XRD patterns of MNO collected at the Elettra synchrotron up to 26.5 GPa are presented in Fig. 8. With increasing pressure, the expected lattice contraction is evidenced by the systematic shift of Bragg reflections toward higher diffraction angles. All the XRD patterns up to 12 GPa can be satisfactorily indexed and refined within the ambient \textit{P-3c1} phase, indicating the structural stability of this phase in the low-pressure regime. Representative Rietveld refinements for the ambient pressure and 11.2 GPa datasets are displayed in Figs. 9(a) and 9(b), respectively. The corresponding structural parameters obtained from the ambient pressure refinement are summarized in Table S1 \cite{SM}. Around 12.8 GPa, a new Bragg reflection emerges near 2$\theta$ = $16.3^\circ$, as indicated by the red arrow in Fig. 8. Concurrently, the (214) reflection around $18.55^\circ$ exhibits pronounced asymmetric broadening, which can be best fitted with three peak components (Fig. S4 ) \cite{SM}, signifying the appearance of two additional reflections. Similarly, a weak new peak begins to develop on the right-hand shoulder of the (104) reflection at 16.4 GPa. Three additional low-angle reflections appear distinctly at 2$\theta$ = 5.8, 6.2, and $6.6^\circ$ at 20.6 GPa, coinciding with the strong emergence of the octahedral Raman mode $\omega$(21). A splitting of the reflection near $8.3^\circ$ is also detected at this pressure. Further new reflections are observed at 23.1 GPa, indicated by upward navy arrows in Fig. 8, along with an additional peak appearing at 24.9 GPa. Two more reflections begin to develop on the right-hand side of the (110) Bragg peak at 23.1 GPa. It is noteworthy that a pronounced transformation in the Raman spectrum, from the ambient \textit{P-3c1} phase to the high-pressure phase is observed around the same pressure (~23 GPa). 

\begin{figure*}[ht!]
\includegraphics[trim={1cm 4cm 3cm 1cm},width=18cm]{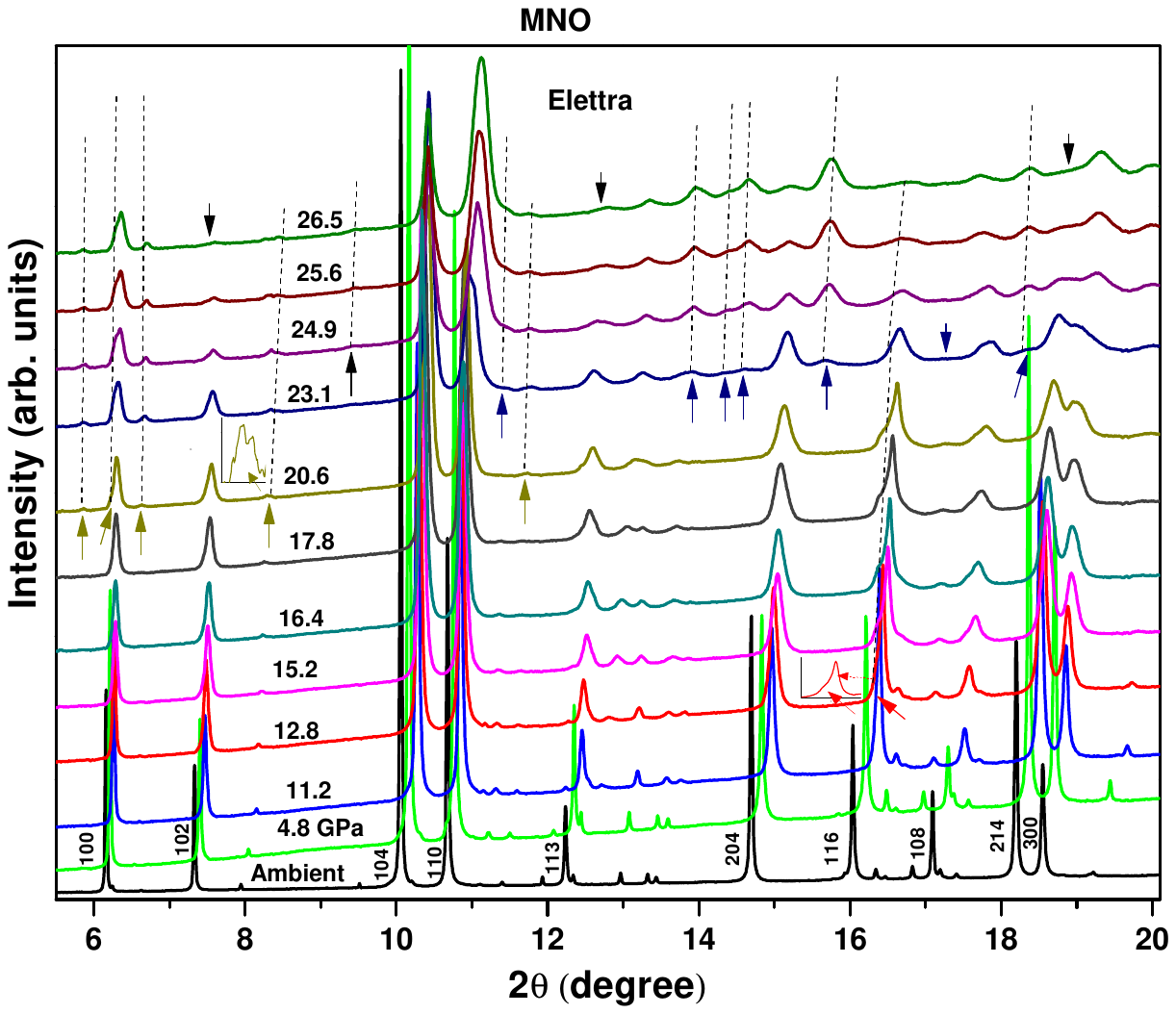}
\caption{\label{fig5} High-pressure XRD patterns of MNO collected at Elettra synchrotron up to 26.5 GPa. The appearance of new Bragg peaks is indicated by upward arrows and traced by the black dashed lines, while the disappearance of peaks is highlighted by downward arrows.} 
\end{figure*}

Indexing of the new reflections at 23.1 GPa yields the best match with a monoclinic \textit{P2/c} symmetry, with lattice parameters comparable to those of the related compound Co$_4$Ta$_2$O$_9$. Rietveld refinement confirms the coexistence of the \textit{P-3c1} and \textit{P2/c} phases in the pressure range of 12.8–26.5 GPa, as illustrated by the refined XRD patterns at 23.1 and 26.5 GPa in Fig. 9(c)-(d). The structural parameters determined at 26.5 GPa are summarized in Table S2 of the SM \cite{SM}. At 26.5 GPa, a few \textit{P-3c1} reflections vanish (downward arrows in Fig. 8(a)), and the prominence feature of newly emerged  peaks is consistent with the stabilization of the dominant \textit{P2/c} structure. However, a minor fraction of the \textit{P-3c1} phase persists even at this pressure, as indicated by Rietveld refinement and corroborated by the weak intensity of A$_{1g}$(4) and A$_{1g}$(5) modes, characteristic of the ambient \textit{P-3c1} phase. The consistent emergence of new Bragg reflections in XRD and the concurrent activation of additional Raman modes around ~13 GPa collectively confirm a pressure-induced structural transformation in MNO, involving the coexistence and gradual dominance of the high-pressure monoclinic \textit{P2/c} phase over the ambient trigonal \textit{P-3c1} structure.

\begin{figure*}[ht!]
\includegraphics[trim={2cm 5cm 4cm 1cm},width=8cm]{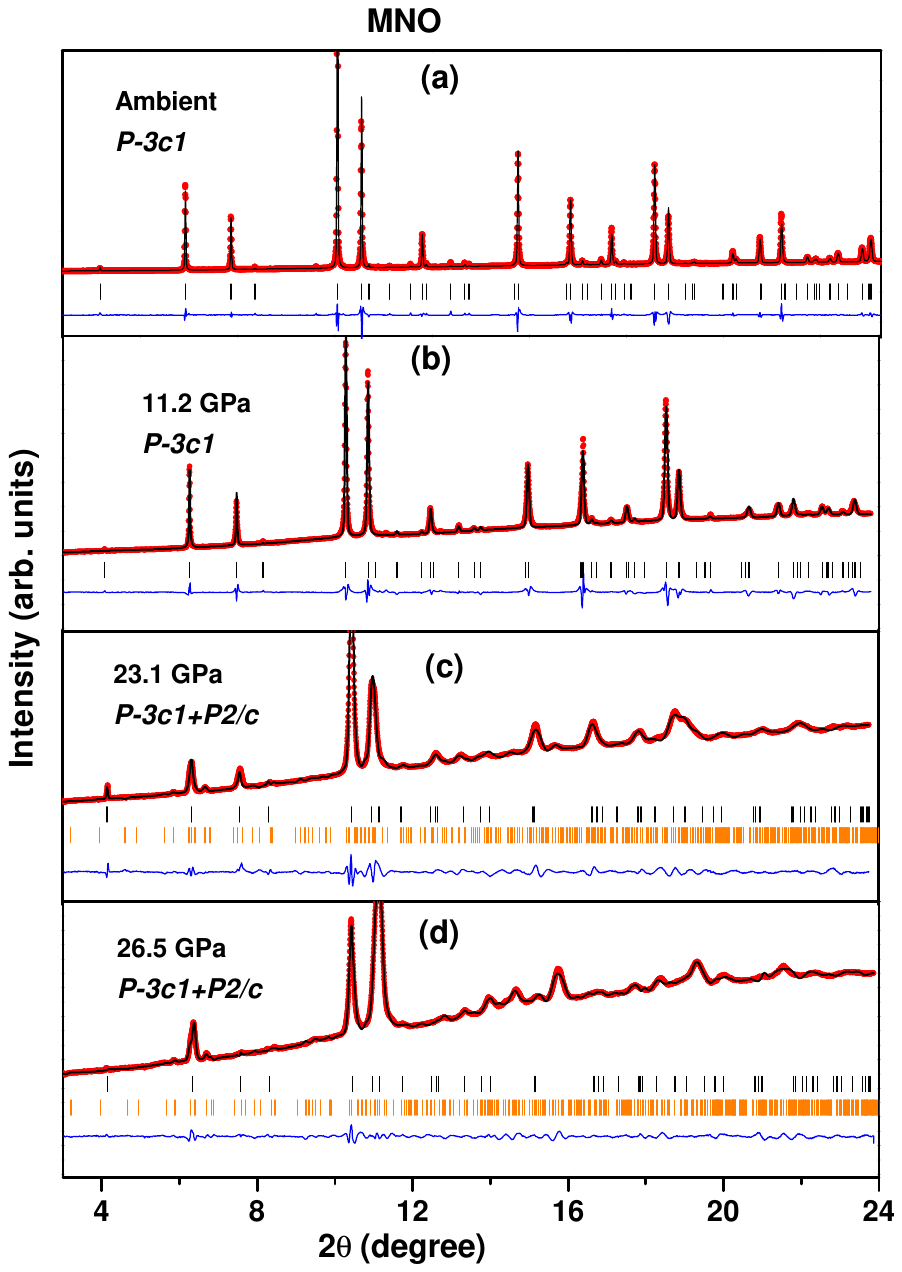}
\caption{\label{fig8} Rietveld-refined XRD patterns at different pressures. Panels (a) and (b) show the \textit{P-3c1} phase at ambient pressure and 11.2 GPa, respectively. Panels (c) and (d) show the coexistence of the \textit{P-3c1} and \textit{P2/c} phases at 23.1 GPa and 26.5 GPa. Red circles represent the observed data, and the solid black line denotes the Rietveld fit. Vertical ticks indicate the calculated Bragg positions for the \textit{P-3c1} (black) and \textit{P2/c} (orange) phases, while the blue bottom line represents the difference between the experimental and calculated intensities. The contribution from gasket is masked by gray shaded region.} 
\end{figure*}

In order to probe the unit-cell modifications around the pressures where Raman modes exhibit anomalies, the pressure evolution of the lattice parameters \textit{a}, \textit{c}, and \textit{c/a} of the \textit{P-3c1} phase, along with the unit-cell volumes of both \textit{P-3c1} and \textit{P2/c} phases, are presented in Fig. 10. For a better understanding of the compressional behavior, the unit-cell volume ($V$) versus pressure ($P$) data are fitted to the third-order Birch–Murnaghan (BM) equation of state (EoS), expressed as \cite{Birch, Murnaghan}:

\begin{equation}
P = {\frac{3}{2}}K_{0}\left[\left(\frac{V_0}{V}\right)^{\frac{7}{3}} - \left(\frac{V_0}{V}\right)^{\frac{5}{3}}\right] \\ \times\left[1+\frac{3}{4}(K'-4)\left[\left(\frac{V_0}{V}\right)^{\frac{2}{3}}-1\right]\right].
\end{equation}

\begin{figure*}[ht!]
\includegraphics[trim={1cm 4cm 3cm 1cm},width=16cm]{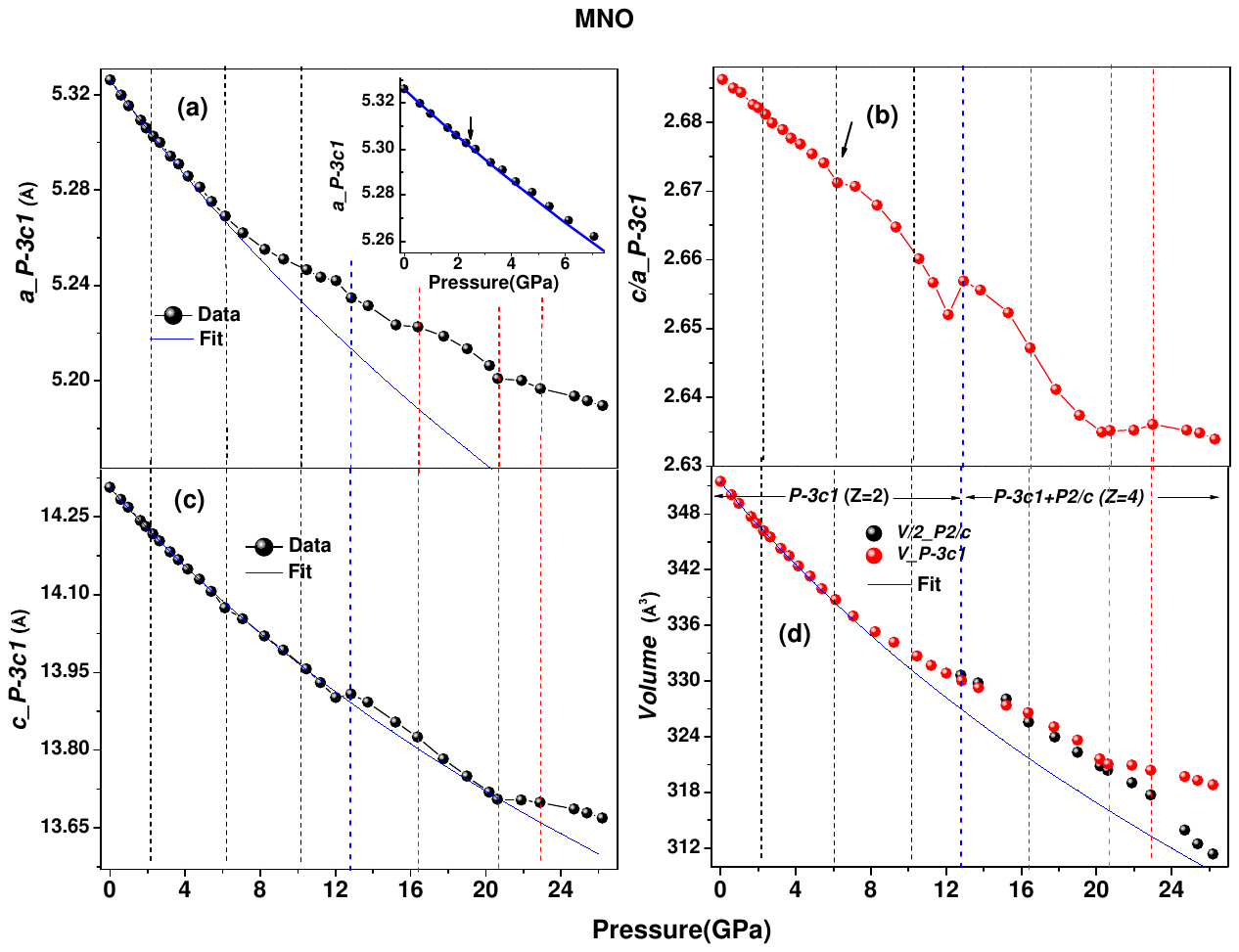}
\caption{\label{fig9} Pressure evolution of MNO unit cell parameters: (a) lattice parameter $a$, (b) $c/a$ ratio, and (c) lattice parameter $c$ for the \textit{P-3c1} phase; (d) unit cell volume for both and phases \textit{P-3c1} and \textit{P2/c}. The solid blue curves represent the Equation of State (EoS) fits to the data using Eq. 1. Vertical markers indicate structural milestones: black dashed lines denote isostructural transition pressures, the blue dashed line highlights the onset of long-range structural transitions, and red dashed lines mark major structural rearrangements at higher pressures.} 
\end{figure*}

where $V_0$ and $K_0$ denote the ambient-pressure volume and bulk modulus, respectively, and $K'$ is the first pressure derivative of the bulk modulus. The same equation is applied to the $a$ and $c$ lattice parameters by replacing $V$ with the cube of the respective parameter.

The compression along the $a$-axis follows the third-order BM EoS well up to ~2 GPa, beyond which it gradually deviates from the fitted curve and shows a pronounced departure above ~6 GPa (Fig. 10(a)). In contrast, the $c$ lattice parameter compresses smoothly up to 12 GPa following the EoS, with a slight discontinuity near 6 GPa (Fig. 10(c)). This discontinuity is more prominently reflected in the \textit{c/a} ratio, as indicated by the black arrow in Fig. 10(b). Consequently, the deviation from the fitted EoS in the unit-cell compression around 6 GPa (Fig. 10(d)) indicates a second isostructural transition, suggesting notable rearrangement of the lattice, consistent with the pronounced slope changes observed in the lower-frequency Raman modes. 

At higher pressures, distinct slope changes are observed in the $a$ parameter around 12.8, 16.2, and 21 GPa, whereas both $c$ and \textit{c/a} display pronounced slope variations near 21 GPa. This behavior correlates with the appearance of the octahedral Raman mode $\omega$(21)—the most intense mode of the \textit{P2/c} phase—and the emergence of several new Bragg reflections around 21 GPa, marking a noticeable structural transformation. The \textit{c/a} ratio of the \textit{P-3c1} phase, which remains nearly constant between 21 and 23 GPa, exhibits a clear downward trend above 23 GPa (Fig. 9(b)). Beyond this pressure, the unit cell volume of the \textit{P2/c} phase shows a markedly higher compression rate, marking a major structural reorganization. This transition is clearly evidenced by the substantial evolution of both the XRD patterns and Raman spectra.

The estimated bulk modulus and its pressure derivative for the unit cell are $K_{0}$ = 145.86(2) GPa, $K'$ = 5.45(1); and those for individual lattice parameters $a$ and $c$ are $K_a$ = 166.32(4) GPa, $K'_a$ = 5.80(3); and $K_{c}$ = 115.52(5) GPa, $K'_c$ = 5.15(2). The obtained unit-cell bulk modulus is lower than those reported for the Nb-based compounds CNO (158.17 GPa) and FNO (155 GPa), and significantly lower than the Ta-based analogue CTO (170.63 GPa) \cite{Jana2, Sahu, Jana3}. The calculated linear compressibilities, $\beta_a$ = 0.006013 GPa$^{-1}$ and $\beta_c$ = 0.008653 GPa$^{-1}$, indicate that the system is ~44\% more compressible along the $c$-axis compared to the $a$-axis. Similar anisotropic compression has been reported in Co-based systems; however, both CNO and CTO show ~20.5\% greater compression along the $c$-axis \cite{Jana2, Jana3}. The remarkably higher anisotropy in MNO may therefore induce stronger modifications in its vibrational, electronic, and magnetic properties under pressure.

	In order to investigate subtle changes associated with isostructural transitions, the Birch–Murnaghan equation of state is linearized in terms of Eulerian strain \cite{Vilaplana, Polian, Dubey}: 
	
			\begin{figure*}[ht!]
\includegraphics[trim={1cm 4.5cm 2cm 1cm},width=12cm]{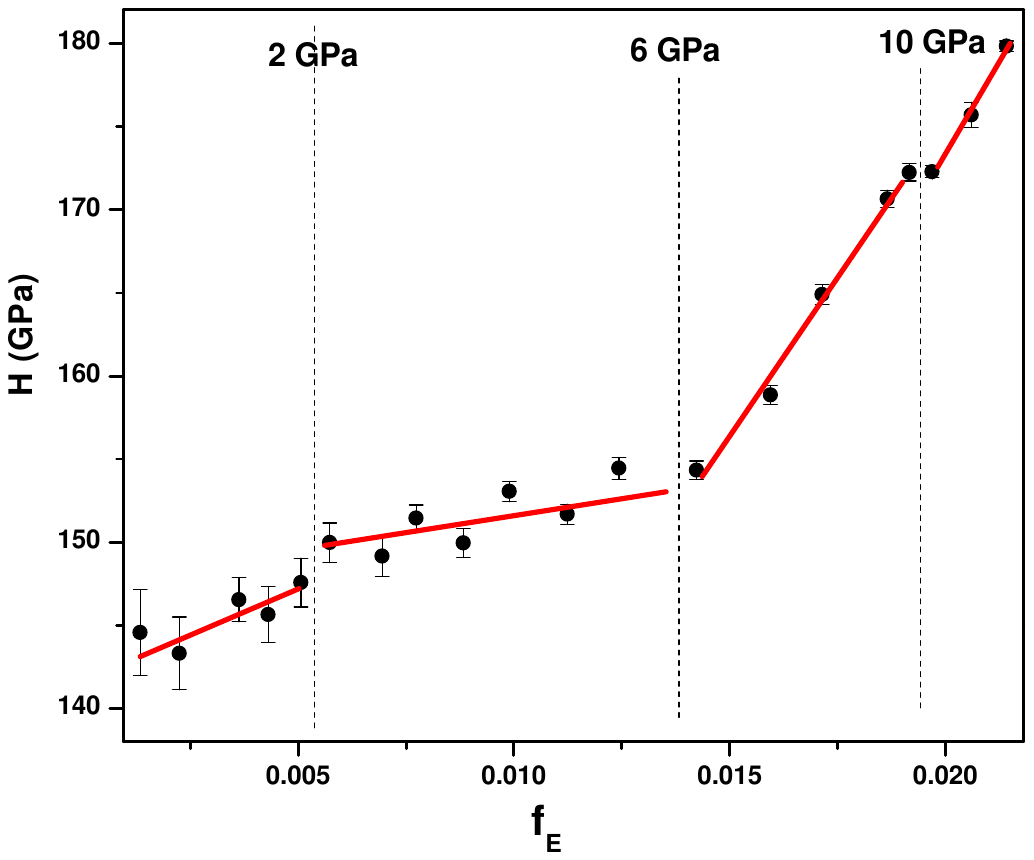}
\caption{\label{fig11} Reduced pressure ($H$) as a function of Eulerian strain ($f_E$) for the trigonal phase of MNO up to 12 GPa. The black dashed lines indicate discontinuities in the $H$-$f_E$ plot, highlighting isostructural phase transitions.} 
\end{figure*}
		
	\begin{equation}
	H = K_0 + \frac{3}{2}K_0(K'-4)f_E,
	\end{equation}
	
	where reduced pressure H is defined as
	
	\begin{equation}
	H = \frac{P}{3f_E(1+2f_E)^{\frac{5}{2}}},
	\end{equation}
	
	and the Eulerian strain f$_E$ is given by 
	
	\begin{equation}
	f_E = \frac{1}{2}\left[\left(\frac{V_0}{V}\right)^{\frac{2}{3}}-1\right]
	\end{equation}

Figure 11 presents the reduced pressure-Eulerian strain in \textit{P-3c1} phase up to 12 GPa. The error bar in $H$ is calulated using the method as described in \cite{Heinz}. In the absence of a structural phase transition, the $H$-$f_E$ plot is expected to follow a linear trend \cite{Polian, Jana6}. However, the $H$-$f_E$ plot of MNO exhibits distinct deviations from linearity, with pronounced slope changes around 2 and 6 GPa, and a subtle discontinuity near 10 GPa. Since no crystal symmetry change is detected within this pressure range, these anomalies likely originate from isostructural and electronic transitions within the system \cite{Vilaplana, Polian, Jana6}.

	\subsection{High pressure synchrotron XRD studies on MTO}

	High-pressure XRD patterns of MTO collected up to 24.8 GPa at the Indus-2 synchrotron are presented in Fig. 12. All diffraction peaks up to 13.1 GPa display the expected lattice compression behavior, characterized by a systematic shift of reflections toward higher diffraction angles. Accordingly, the XRD data within this pressure range are well fitted using the ambient \textit{P-3c1} structural model as shown for the ambient and 7.1 GPa patterns in Fig. S5 \cite{SM}. The structural coordinates obtained from the ambient-pressure Elettra data are presented in Table S1 \cite{SM}. A new Bragg reflection emerges on the left-hand side of the (100) peak at 15 GPa, as indicated by the red arrow in Fig. 12. This feature closely resembles the behavior observed in MNO at 20.6 GPa, where a significant transformation from the \textit{P-3c1} to \textit{P2/c} phase occurs. Around 18 GPa, pronounced asymmetric broadening of the (110) peak is evident, signaling the development of an additional peak on its right-hand shoulder. Upon further compression to 20.4 GPa, another distinct reflection appears on the right side of the (100) peak, accompanied by two additional high-angle reflections.

\begin{figure*}[ht!]
\includegraphics[trim={1cm 4cm 2cm 0cm},width=14cm]{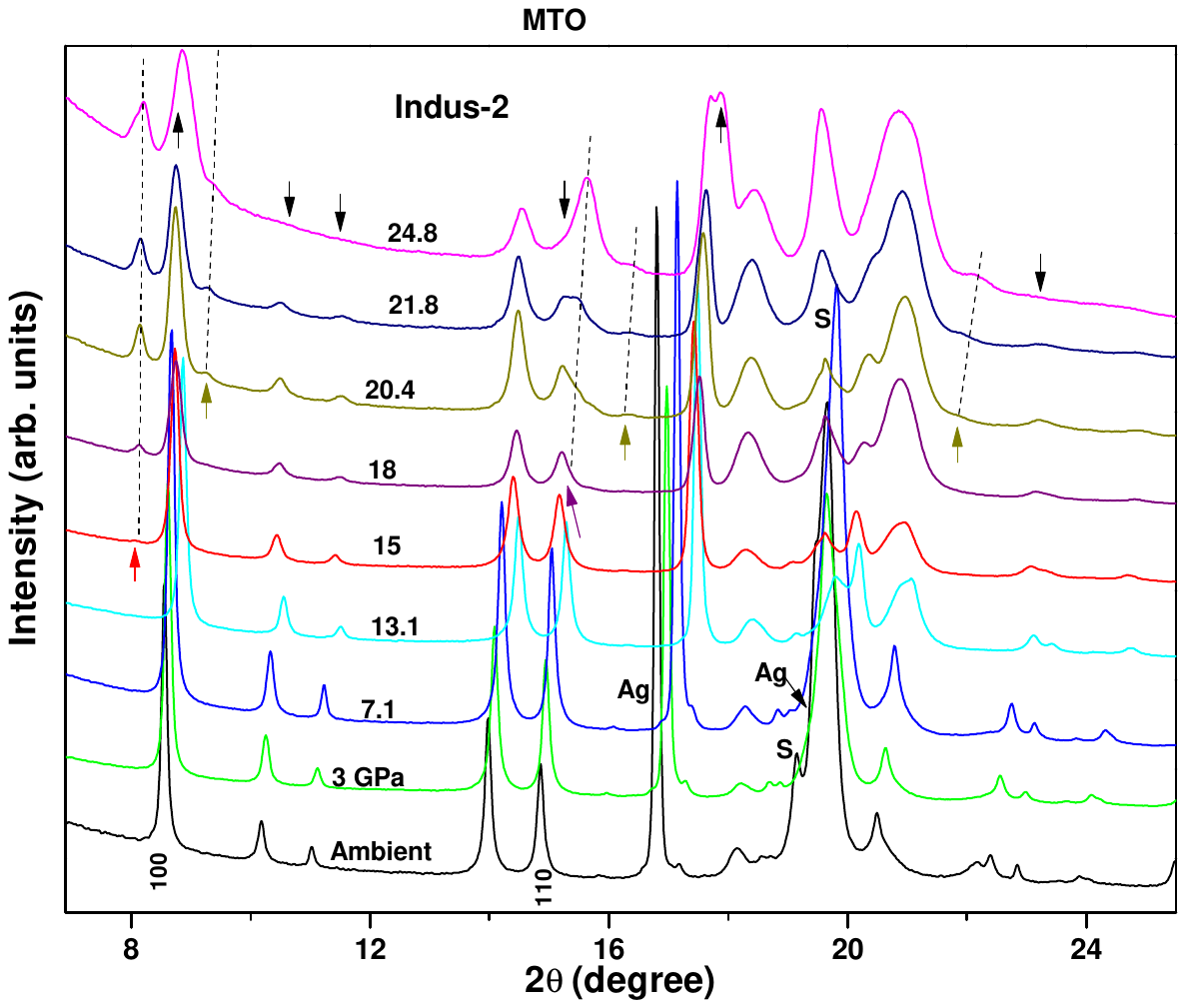}
\caption{\label{fig1} High-pressure XRD patterns of MTO collected at the Indus-2 synchrotron source up to 24.8 GPa. The appearance and disappearance of Bragg reflections are denoted by upward and downward arrows, respectively. Black dashed lines trace the pressure evolution of emergent peaks. Peaks labeled Ag and S correspond to the silver pressure marker and the gasket material, respectively.} 
\end{figure*}

	At the highest pressure investigated (24.8 GPa), several \textit{P-3c1} reflections vanish, while two new diffraction lines emerge, suggesting substantial structural reorganization. Overall, the trend and positions of these newly emerged reflections are comparable to those observed in the Nb analogue MNO, suggesting a similar pressure-driven structural evolution toward a monoclinic distortion. However, fewer new peaks are discernible in MTO, primarily due to the lower brilliance and broader diffraction profiles of the Indus-2 synchrotron beam. Consequently, the high-pressure structural analysis of MTO is performed using the same \textit{P-3c1} + \textit{P2/c} mixed-phase model as employed for MNO. The diffraction data between 15 and 21.8 GPa can be satisfactorily indexed using a combination of \textit{P-3c1} and \textit{P2/c} symmetries, as demonstrated by the representative refinement at 21.8 GPa in Fig. S5 \cite{SM}, with the corresponding structural parameters provided in Table S3 \cite{SM}. This indicates a gradual pressure-induced structural transformation from the ambient \textit{P-3c1} phase. At 24.8 GPa, the Le Bail profile fitting is employed due to the pronounced peak broadening in the diffraction data. The analysis reveals that the pattern can be reasonably fitted with a single \textit{P2/c} phase, suggesting the near-complete stabilization of the high-pressure monoclinic structure. However, given the broad diffraction features, the persistence of a minor  \textit{P-3c1} phase at this pressure cannot be entirely excluded.

The evolution of the lattice parameters for the \textit{P-3c1} phase, along with the unit cell volume of the \textit{P-3c1} and \textit{P2/c} phase, is presented in Fig. 13. The equation of state (EoS) fitting for the unit cell volume and lattice parameters is performed using Eq. (1), as described in the preceding section. The pressure evolution of the $a$-lattice parameter follows the Birch–Murnaghan EoS up to 10 GPa, beyond which a slight deviation becomes evident (Fig. 13(a)). Around 15 GPa, the compression along the $a$ axis exhibits a pronounced deviation toward higher values relative to the extrapolated fit. While, the $c$-lattice parameter shows a subtle deviation from the BM EoS above 6 GPa, which increases progressively beyond 10 GPa and reaches a maximum deviation near 15 GPa—coinciding with the onset of the monoclinic \textit{P2/c} phase. At higher pressures, the compression of the $c$ parameter returns closer to the EoS trajectory.

\begin{figure*}[ht!]
\includegraphics[trim={1cm 4.5cm 3cm 1cm},width=16cm]{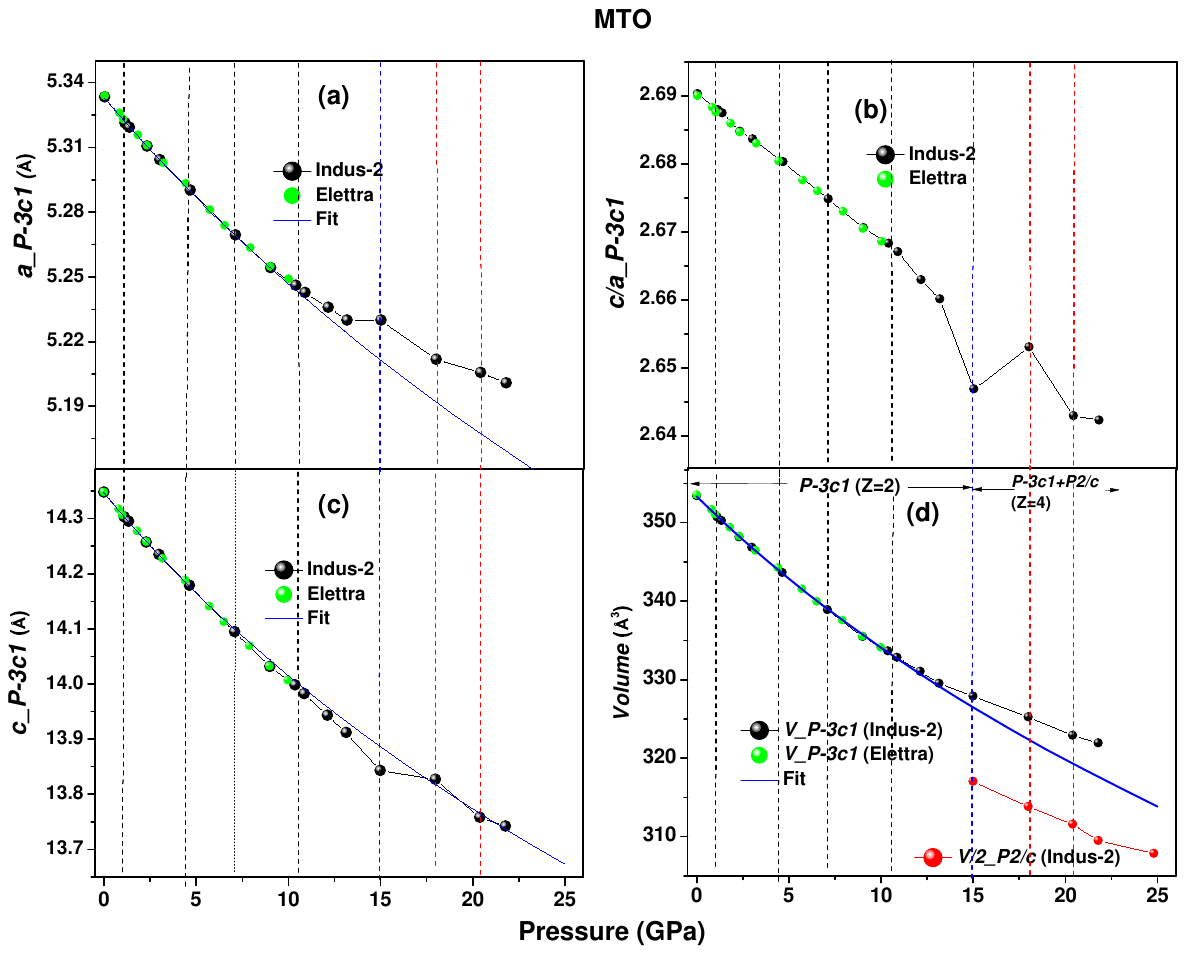}
\caption{\label{fig11}(Colour online) Pressure evolution of MTO unit cell parameters obtained from the Elettra (up to 9 GPa) and Indus-2 (up to 24.8 GPa) synchrotrons. (a)-(c) Lattice parameters for the phase \textit{P-3c1}; and (d) unit cell volume for the \textit{P-3c1} and $P2/c$ phases. Solid blue curves denote EoS fits. Distinct structural anomalies are highlighted by vertical indicators: black dashed lines for isostructural transitions, a blue dashed line for the initiation of long-range structural transitions, and red dashed lines identifying significant lattice rearrangements at higher pressures.} 
\end{figure*}

	The irregularities observed around 10 GPa in the $a$-axis and at 6 and 10 GPa in the $c$-axis likely correspond to isostructural transitions, consistent with the Raman anomalies detected near these pressures. The maximum deviations in both $a$ and $c$ parameters at 15 GPa signify substantial lattice distortions associated with the initiation of a long-range structural transformation. This transition is also reflected in the distinct drop in the \textit{c/a} ratio at 15 GPa. Furthermore, additional slope changes in the \textit{c/a} ratio are observed around 18.1 GPa, and 20.4 GPa, aligning with the pressures at which anomalies appear in several Raman mode frequencies and linewidths.

	The bulk modulus estimated from the 3$^{rd}$-order BM equation of state for the unit cell volume and the lattice parameters  $a$ and $c$ of MTO are 153.29(2), 177.29(3), and 118.57(1) GPa, respectively. The pressure derivatives are determined to be 5.38(4), 6.25(2), and 5.23(1) for the volume, $a$-axis, and $c$-axis, respectively. Table I provides a comparative overview of the EoS parameters for MNO, MTO, and their Co-based analogues. The results indicate that Mn-based systems are generally more compressible than their Co counterparts, as reflected in the lower bulk moduli. Furthermore, the $B$-site variation from Nb to Ta reduces compressibility in both Mn- and Co-based compounds. The disparity between the $a$- and $c$-axis compressibilities is significantly more pronounced in the Mn-based systems, resulting in strongly anisotropic compression along the $c$ axis. This pronounced anisotropy is consistent with the pressure-dependent behavior observed in Raman measurements.

			\begingroup
 \squeezetable	
		\begin{table*}[htb]
		 \squeezetable
\caption{\label{tab:table2
}
Comparison of the Equation of State (EoS) parameters for the unit cell and lattice constants of MNO, MTO, and their Co-based analogues \cite{Jana2, Jana3} in the phase \textit{P-3c1}.}
\begin{ruledtabular}
\begin{tabular}{ccccccccccccc}
 Material  & $V_0$ (\AA$^3$) & $K_{0}$ (GPa) & $K'$  &  $K_a$ (GPa)  & $K_c$ (GPa)  &  $\beta_a$ &  $\beta_c$ & anisotropy (\%) \\
\hline
 CNO & 326.71(2) & 158.17(3) & 5.34(5) & 163.62(3)     & 135.59(1)   & 0.00204  & 0.00246 & 20.5 \\
\hline
 CTO & 327.79(1)   & 170.63(5) &  4.97 (1)   &  183.57(5)     & 152.47(6)   & 0.001815 & 0.002186 & 20.4   \\
\hline 
 MNO & 351.49(3) & 145.86(2)      & 5.45(1)    & 166.32(4)        & 115.52(5)     & 0.006012 & 0.008567 & 42.5 \\
\hline
 MTO & 353.34(3)  & 153.29(2)      & 5.38(4) & 177.29(3)         & 118.57(1)     & 0.005641 & 0.008418 & 49.2 \\

\end{tabular}
\end{ruledtabular}
\end{table*}

Figure 14 presents the reduced pressure  versus Eulerian strain profile of MTO for the \textit{P-3c1} phase up to 13 GPa. In the initial pressure range of 0.6–1 GPa, the reduced pressure exhibits an unusual decrease, indicative of structural or electronic instability within the lattice. This anomaly aligns well with the local symmetry breaking detected at 0.5 GPa in Raman measurements and the redshift of low-frequency Raman modes up to 0.8 GPa, both of which suggest an early pressure-induced lattice distortion or electronic rearrangement. Above 1 GPa, the reduced pressure shows a steep increase up to 3.2 GPa, followed by a downward trend in the 3.2–6 GPa range. Between 6 and 10 GPa, the $H$–$f_E$ relationship remains nearly constant, before gradually rising again up to 13 GPa. These distinct irregularities around 1, 3.2, 6, and 10 GPa coincide closely with the anomalous behavior observed in the Raman mode frequencies and linewidths. In the absence of any average symmetry change, such pronounced slope variations in the $H$–$f_E$ plot are strong signatures of isostructural and electronic transitions in the system, reflecting subtle modifications in bond compressibility and pressure-induced changes in electronic/magnetic properties.

	\begin{figure*}[ht!]
\includegraphics[trim={1cm 4.5cm 2cm 1cm},width=14cm]{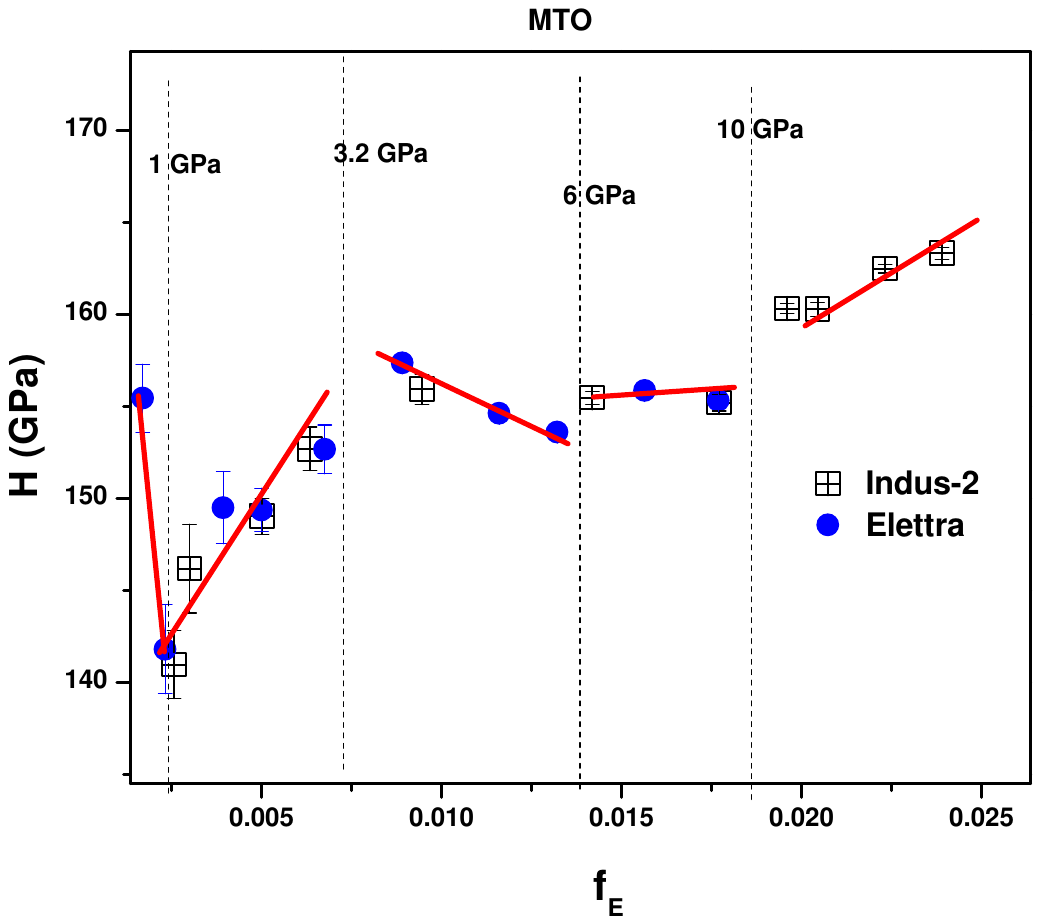}
\caption{\label{fig14} Reduced pressure–Eulerian strain ($H$-$f_E$) plot for MTO in \textit{P-3c1} phase up to 13 GPa. The black dashed lines indicate irregularities in the linear behavior, signifying the occurrence of isostructural transitions.} 
\end{figure*}


\subsection{DFT-based computations}
	At ambient pressure, full structural optimization is performed for both MNO and MTO within the trigonal \textit{P-3c1} space group. The calculated lattice parameters align well with experimental parameters, confirming the stability of the trigonal arrangement under ambient conditions. To investigate pressure-induced structural evolution, we evaluated enthalpy as a function of external pressure for both the ambient trigonal \textit{P-3c1} and high-pressure monoclinic \textit{P2/c} phases. For MNO, the enthalpy curves in Fig. 15(a) show the monoclinic phase approaching the trigonal phase at approximately 11 GPa, becoming energetically favorable near 27.5 GPa. This behavior suggests the onset of a structural transformation around 11 GPa, where the two phases become nearly degenerate. The complete stabilization of the monoclinic phase beyond $\sim$27.5 GPa indicates a broad coexistence or mixed-phase regime, consistent with experimental findings. A similar trend is observed for MTO. As shown in Fig. 15(b), the enthalpy of the phase approaches that of the phase near 14 GPa and becomes more favorable above approximately 27 GPa. These results indicate that MTO undergoes a comparable pressure-induced transition with a similar coexistence region. The enthalpy crossover between these competing phases provides strong theoretical support for a pressure-driven symmetry reduction.
    
    	\begin{figure*}[ht!]
\includegraphics[trim={1cm 1.5cm 1cm 1cm},width=15cm]{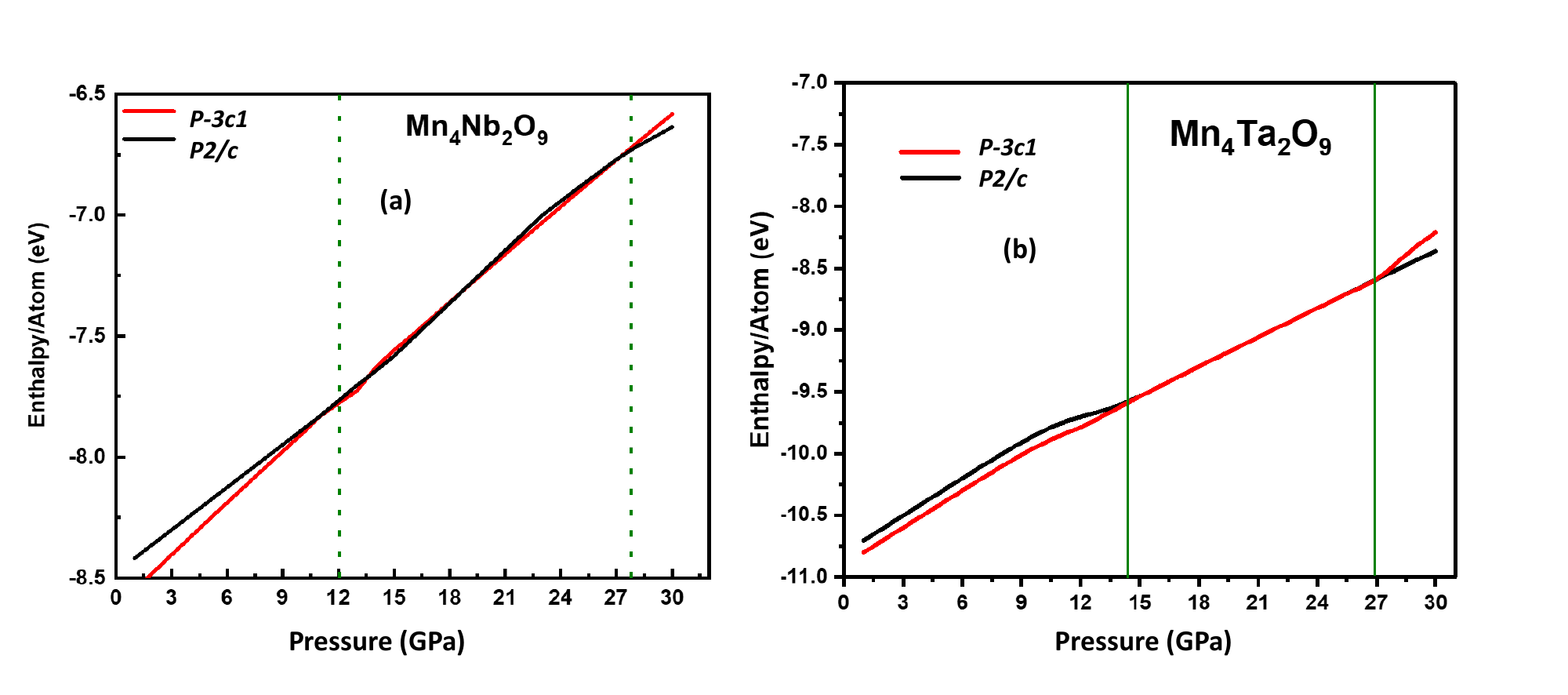}
\caption{\label{fig15} Pressure-dependent enthalpy variations for the trigonal \textit{P-3c1} and monoclinic \textit{P2/c} phases of (a) MNO and (b) MTO. Dashed lines indicate the onset of long-range structural transitions and the subsequent stabilization of the monoclinic phase.} 
\end{figure*}

\subsection{Discussion}


Within the linear magnetoelectric $A_4B_2$O$_9$ family, only the Fe-based compound Fe$_4$Nb$_2$O$_9$ exhibits a long-range structural transition from trigonal \textit{P-3c1} to monoclinic \textit{C2/c} below its N\'eel temperature \cite{Jana}. An identical transition has also been reported under pressure near 9 GPa \cite{Sahu}. In the other Fe-based analogue, Fe$_4$Ta$_2$O$_9$, no such distortion has been observed in laboratory-based x-ray diffraction measurements at low temperatures, and high-pressure studies are yet to be reported. Moreover, neutron diffraction or synchrotron-based x-ray diffraction experiments, which are more sensitive to subtle oxygen-driven distortions are still lacking for FTO. In contrast, pressure-induced multiple isostructural and average crystallographic transitions have been reported in the Co-based variants. CNO exhibits three isostructural transitions at 5.2, 8.5, and 11.5 GPa, accompanied by pronounced Raman anomalies and the emergence of new modes resembling its low-temperature behavior associated with short-range magnetic ordering, suggesting the possible development of magnetic interactions under compression \cite{Jana2}. Similar transitions are also observed in CTO, but at significantly lower pressures (3.0, 5.5, and 7.7 GPa), which has been attributed to the stronger spin–orbit coupling in the Ta-based compound \cite{Jana3}. At higher pressures, CNO evolves into a mixed \textit{C2/c} + \textit{P2$_1$/c} phase above 21.5 GPa, via an intermediate \textit{P-3c1} + \textit{C2/c} coexistence region between 17 and 20 GPa \cite{Jana2}. In contrast, CTO follows a distinct transition sequence, partially transforming into a more distorted \textit{P2/c} phase at $\sim$13 GPa, followed by the emergence of an additional \textit{C2/c} phase above 18.4 GPa \cite{Jana3}. In the present study, three isostructural transitions are identified in MNO at approximately 2.0, 6.6, and 10 GPa, while four such transitions are detected in MTO at 0.5, 3.2, 6.0, and 10 GPa. Thus, similar to the Co-based systems, the Ta-based Mn compound undergoes isostructural distortions at lower pressures than its Nb analogue. However, MNO transforms into a partially monoclinic phase at a slightly lower pressure ($\sim$12.5 GPa) than MTO ($\sim$14 GPa), which contrasts with the trend observed in the Co-based compounds, where the Ta analogue exhibits the first long-range symmetry transition at substantially lower pressure (13 GPa) than the Nb counterpart (17 GPa).

Furthermore, distinct differences in the stabilization of high-pressure symmetry are observed in MNO relative to its Co- and Fe-based analogues. While both FNO and CNO undergo a direct transition to the monoclinic \textit{C2/c} phase at the onset of their first long-range structural transformation, MNO instead partially transforms into a \textit{P2/c} phase. This contrasting behavior is likely related to differences in the local structural environment among these compounds. Raman spectroscopy reveals that MNO is significantly more locally distorted than CNO already at ambient conditions \cite{Jana4}. Consistently, nuclear magnetic resonance studies show that the local structure of MNO differs substantially from that of CNO, while exhibiting closer similarity to its Ni-based analogue NNO \cite{Jana5}, despite the stabilization of different average crystal symmetries in these systems. High-pressure investigations on NNO further support this interpretation, as a long-range structural transition from orthorhombic \textit{Pbcn} to a partially transformed \textit{P2/c} phase—its direct subgroup—is observed at $\sim$12.7 GPa \cite{Jana5}, remarkably close to the transition pressure in MNO. Notably, NNO also exhibits three isostructural transitions at approximately 2.0, 6.2, and 10 GPa, nearly identical to those observed in MNO. In addition, both systems undergo substantial transformation into a \textit{P2/c}-dominated phase in the pressure range of 22–23 GPa, as evidenced by the pronounced evolution of their Raman spectra. These striking similarities further highlight the close local structural analogy between MNO and NNO \cite{Jana5}. MTO also exhibits a similar average structural transition and a \textit{P2/c}-dominated high-pressure phase around 22–23 GPa, comparable to MNO and NNO, consistent with the analogous local structures of MNO and MTO \cite{Jana4}. Analysis of the ambient-pressure Raman spectra indicates that, while CTO is considerably more locally distorted than CNO, the two Mn-based systems, MNO and MTO, possess closely related local structural environments compared to their Co-based counterparts \cite{Jana4}. Nevertheless, their Raman spectra are not identical and display several distinct features, which may account for their differing responses during the low-pressure isostructural transitions.
These differences may originate from variations in orbital hybridization involving Nb $4d$ and Ta $5d$ electrons with the $A$-site cation orbitals and oxygen $2p$ states. Additionally, the stronger spin–orbit coupling in the heavier Ta-based variant compared to its Nb counterpart is expected to further influence the pressure response of the structural, electronic, and vibrational properties.

Among the Mn- and Co-based systems, MTO exhibits the most pronounced changes in Raman mode frequencies and linewidths, which can be attributed to its strong anisotropic lattice compression under pressure. Our low-temperature studies have already established robust spin–phonon coupling in these materials, wherein magnetic ordering modifies the lattice through phonon interactions. Conversely, the application of external pressure can alter structural parameters and, in turn, activate magnetic interactions. In MTO, pronounced anomalies in Raman mode frequencies are observed around 0.5 and 3.2 GPa. Notably, the substantial softening of low-frequency modes above 3.2 GPa, relative to the 0.8–3.2 GPa regime, closely mirrors the behavior reported below the short-range magnetic ordering temperature at ambient pressure \cite{Jana4}. Furthermore, the appearance of the $\nu$(2) mode at 0.5 GPa coincides with its activation below the long-range magnetic ordering temperature \cite{Jana4}. In an ideal isostructural transition, no significant linewidth anomalies are expected, as the Wyckoff positions remain unchanged. However, the marked irregularities in mode linewidths observed at 0.5 and 3.2 GPa suggest the onset of local lattice distortions coupled with evolving magnetic correlations. Collectively, these features point toward the reemergence of short-range magnetic order in the 0.5–3.2 GPa pressure range. A similar correspondence is evident in MNO, which displays reproducible anomalies in both mode frequency and linewidth at 2 and 6.6 GPa, consistent with its low-temperature magnetic behavior \cite{Jana4}. The emergence of the $\omega$(2) mode at 6.6 GPa—previously observed only in the magnetically ordered state at 77 K \cite{Jana4}—further indicates the development of pressure-induced short-range magnetic correlations in the 2–6 GPa regime. Comparable signatures of coupled magnetic, vibrational, and structural degrees of freedom have also been reported in the Co-based analogues, occurring at 3–5 GPa for CTO and 5–7 GPa for CNO \cite{Jana2, Jana3}. These results collectively demonstrate that, in the \textit{A}$_4$\textit{B}$_2$O$_9$ linear magnetoelectric family, moderate pressures of only a few gigapascals are sufficient to effectively tune spin–lattice coupling. This sensitivity underscores their strong potential for spintronic and multifunctional device applications, particularly in epitaxial thin-film geometries, where comparable strain-induced pressures are readily attainable. 	

Moreover, indirect evidence for re-entrant magnetic interactions is revealed through the pressure evolution of the linewidths of the $\nu$(1) and $\nu$(2) modes in MTO. These pressure-induced modes display behavior distinctly different from that of the other Raman modes: the linewidth of $\nu$(1) increases by nearly a factor of two in the 0.5–3.8 GPa range, while that of $\nu$(2) exhibits an almost fivefold enhancement between 3.2 and 5.8 GPa. Such pronounced broadening cannot be accounted for solely by structural effects and instead suggests the involvement of additional scattering channels, likely of magnetic origin. To further elucidate this behavior, the pressure dependence of the integrated intensities of these modes, together with other low-frequency Raman modes, is shown in Fig.~16. The intensities of Raman modes associated with the ambient \textit{P-3c1} structure generally decrease with increasing pressure up to 6 GPa, aside from minor irregularities near 0.5 and 3.2 GPa. In contrast, the intensity of the $\nu$(1) mode increases gradually up to 5.8 GPa and exhibits a drop around 6 GPa, followed by a sharper enhancement near the onset of the structural transition. The $\nu$(2) mode exhibits an even more anomalous behavior, with its intensity continuing to increase up to 21 GPa, in stark contrast to the ambient-phase modes. Considering that the $\nu$(2) mode emerges at $T_N$ in our low-temperature Raman measurements \cite{Jana4}, and given its unusual linewidth broadening and intensity enhancement under pressure, its origin may be attributed to magnetically activated scattering processes \cite{Jana7}. Similarly, the appearance of the $\nu$(1) mode near $T_{sro}$ at ambient pressure and its distinct pressure evolution further suggest a magnetic origin. A comparable, albeit less pronounced, behavior is also observed in MNO, where the $\Omega$(1) and $\Omega$(2) modes display pressure dependencies that differ from those of the ambient-phase Raman modes. Interestingly, weakly intense $\nu$(1) and $\nu$(2) modes have also been reported around 3 GPa in the Co-based Ta analogue CTO \cite{Jana3}. In contrast, in CNO the $\nu$(1) mode appears below $T_{sro}$ at ambient pressure and emerges above 5.2 GPa at room temperature, as shown in our earlier work \cite{Jana2}. Notably, in CNO the $\nu$(1) mode also exhibits a distinct pressure evolution of both linewidth and intensity compared to the Raman modes of the ambient phase \cite{Jana2}. Taken together, these consistent observations across Mn- and Co-based compounds highlight the intimate coupling among magnetic, vibrational, and structural degrees of freedom and strongly suggest the presence of pressure-induced re-entrant magnetostructural correlations in the $A_4B_2$O$_9$ family. Among these systems, MTO is particularly remarkable, as it exhibits multiple anomalous signatures beginning at pressures as low as 0.5 GPa—significantly lower than in its analogues—underscoring its exceptional sensitivity to external pressure.

	\begin{figure*}[ht!]
\includegraphics[trim={1cm 6cm 2cm 1cm},width=12cm]{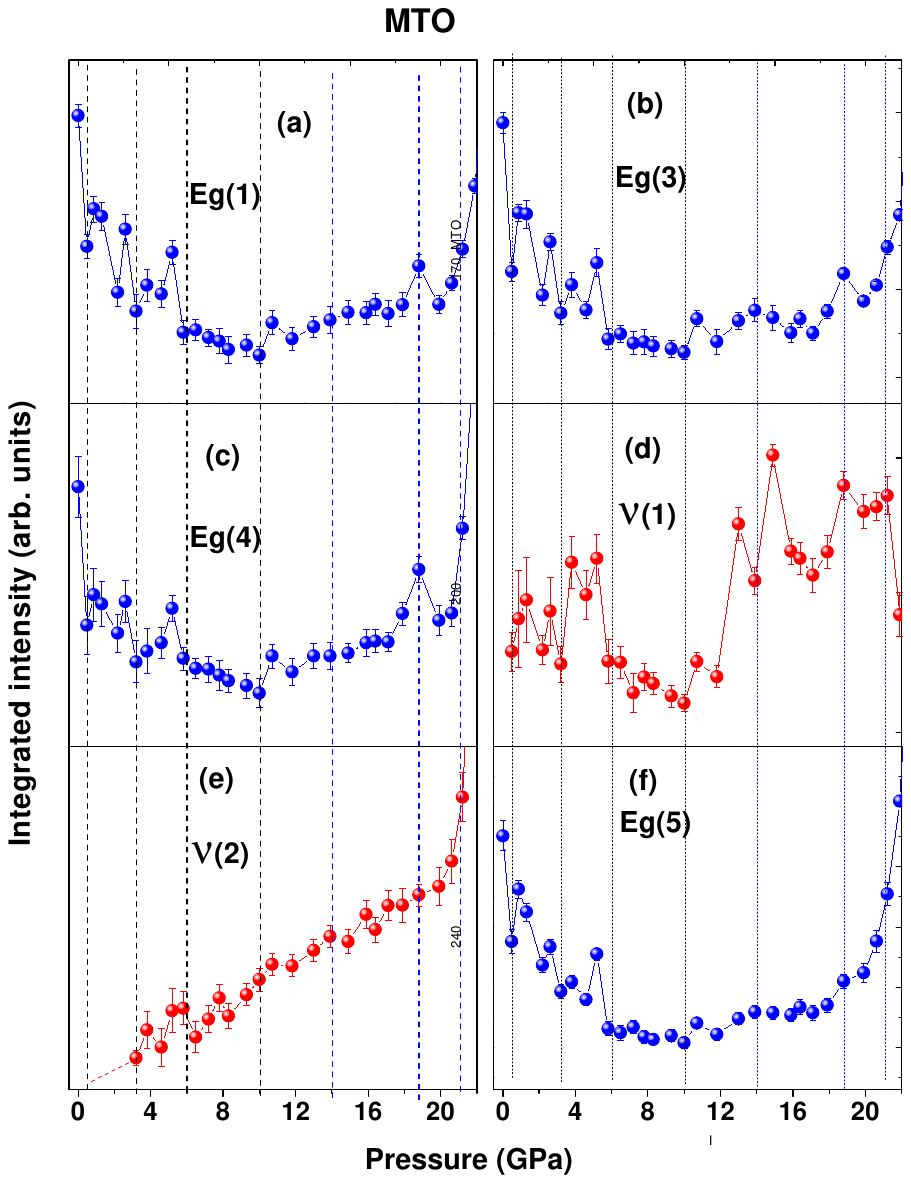}
\caption{\label{fig15}(Colour online) (a)–(f) Pressure dependence of the integrated intensities for selected Raman modes of MTO. Vertical black dashed lines indicate the onset of isostructural transitions, while blue dashed lines denote the pressure thresholds for long-range structural transition and significant lattice rearrangements.} 
\end{figure*}
	
	Pawbake \textit{et al.} reported a dramatic enhancement of interlayer exchange interaction above 6 GPa in the layered van der Waals (vdW) magnet CrSBr, originating from a pressure-induced reduction in the interlayer spacing \cite{Pawbake}. This study further demonstrated that external pressure can directly modify both the magnetic exchange interactions and the magnetocrystalline anisotropy arising from spin–orbit coupling \cite{Pawbake}. A similar pressure-driven enhancement of magnetic ordering has been observed in another vdW magnet, NiI$_2$, where the Néel temperature increases from 73 K at ambient pressure to 190 K at 8.6 GPa due to strengthened interlayer exchange interactions \cite{Liu}. Likewise, anisotropic compression along the $c$ axis leading to the stabilization of a magnetically ordered phase has been reported in the honeycomb-layered vdW material CrCl$_3$ \cite{Lis}.
Magnetocrystalline anisotropy is known to play a crucial role in realizing long-range magnetic order in two-dimensional isotropic Heisenberg systems, where such ordering is otherwise prohibited by the Mermin–Wagner theorem \cite{Gong, Huang}. In the present $A_4B_2$O$_9$ family, a significant contribution from single-ion anisotropy to the magnetic Hamiltonian has been reported \cite{Deng2, Jana4}. External pressure is therefore expected to be an effective tuning parameter for both the Heisenberg exchange interaction and the single-ion anisotropy, as clearly demonstrated in layered CrSBr \cite{Pawbake}. In the present MNO and MTO systems, a 40–49\% higher compressibility along the $c$ axis compared to the $a$ axis results in pronounced anisotropic lattice compression. This renders these materials analogous to vdW-like systems, where weakly bonded layers can be readily manipulated by external pressure, thereby strongly modifying magnetic exchange interactions and single-ion anisotropy. A similar scenario is anticipated for the Co-based analogues CNO and CTO, which exhibit an anisotropic compression of approximately 20.5\% along the $c$ axis. In our recent comparative study of Mn- and Co-based systems, we demonstrated a strong correlation between the magnetic ordering temperature and the $c/a$ ratio \cite{Jana4}. To further examine this relationship, Fig.~17(a) compares the pressure evolution of the $c/a$ ratio for all four Mn- and Co-based compounds, while Fig.~17(b) illustrates the correlation between $T_\mathrm{N}$ and the $c/a$ ratio across various $A$-site and $B$-site compositions and substitutions in the $A_4B_2$O$_9$ family \cite{Maignan, Panja, Schwarz, Panja2, Jana, Yu, Jana2, Jana3, Jana4, Zheng2, Chaudhary, Maignan2}. 

\begin{figure*}[ht!]
\includegraphics[trim={1cm 4cm 2cm 1cm},width=14cm]{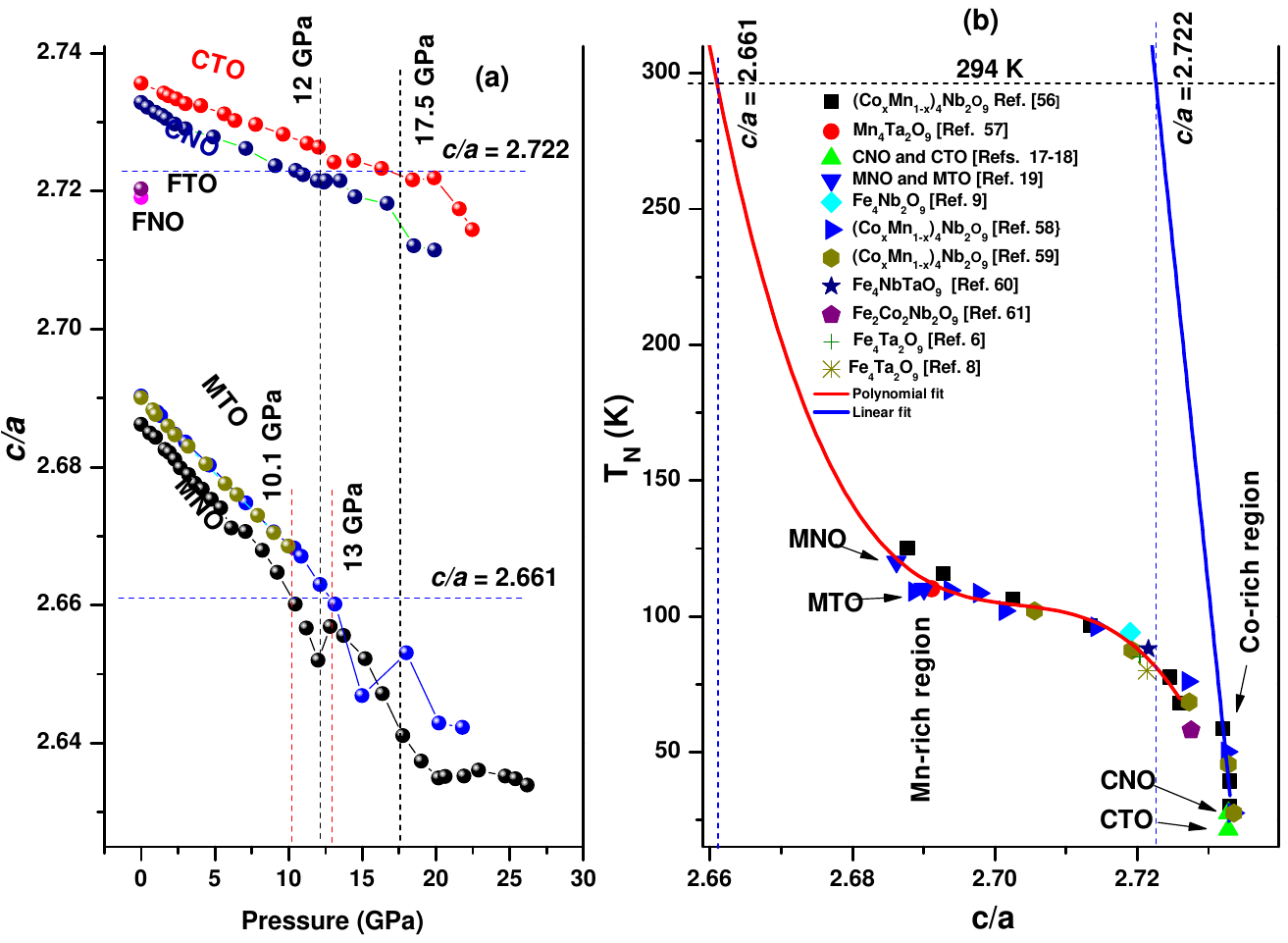}
\caption{\label{fig11} (a) Comparison of the $c/a$ ratio as a function of pressure for MNO, MTO and their Co-bsed analogue \cite{Jana2, Jana3}. (b) Variation of $T_N$ with the \textit{c/a} ratio  across various $A$- and $B$-site compositions within the the $A_4B_2$O$9$ family. The blue dashed line indicates the critical $c/a$ ratio required to reach a $T_N$ of 294 K (room temperature). Red and black dashed lines denote the estimated pressure values required to achieve this ratio for the Mn- and Co-based analogues, respectively.} 
\end{figure*}

Although the magnitude of the $c/a$ reduction is smaller in the Co-based systems than in the Mn-based counterparts, their Néel temperatures are significantly more sensitive to changes in the $c/a$ ratio. This enhanced sensitivity likely originates from the canted spin structure and stronger single-ion anisotropy associated with the larger spin–orbit coupling of Co$^{2+}$ compared to Mn$^{2+}$ \cite{Jana4}. The Co-rich region exhibits a steeper increase in $T_{N}$ with decreasing $c/a$, which can be well described by a linear fit. In contrast, the remaining compositional space is better captured by a third-order polynomial function, revealing a pronounced enhancement of $T_\mathrm{N}$ in the Mn-rich region. Extrapolation of these fits suggests that achieving room-temperature magnetic ordering ($T_\mathrm{N} \approx 294$ K) would require reducing the $c/a$ ratio to approximately 2.661 for Mn-based systems and 2.722 for Co-based systems. Analyzing the pressure dependence of the $c/a$ ratio indicates that pressures of approximately 10.1 GPa for MNO and 13 GPa for MTO would be sufficient to reach the required values. Similarly, pressures of about 12 GPa and 17.5 GPa would be needed for CNO and CTO, respectively. These results strongly suggest that long-range magnetic ordering can be realized in the pressure range of 10–17 GPa, while short-range magnetic correlations can already be induced at much lower pressures, as discussed above.

\section{Conclusion}

Our combined high-pressure Raman spectroscopy, synchrotron x-ray diffraction and DFT calculations reveal a rich sequence of pressure-induced isostructural and long-range structural transitions in the Mn-based honeycomb magnetoelectric compounds Mn$_4$Nb$_2$O$_9$ and Mn$_4$Ta$_2$O$_9$. Both systems exhibit multiple isostructural transitions prior to a partial transformation from the ambient trigonal \textit{P-3c1} phase to a monoclinic \textit{P2/c} phase, with MTO showing the earliest onset of local symmetry breaking at an exceptionally low pressure of 0.5 GPa. The pronounced anisotropic lattice compression, characterized by a substantial reduction in the $c/a$ ratio, underscores the key role of pressure in enhancing interlayer coupling. The emergence of Raman modes analogous to those reported at low temperatures, together with anomalous linewidth and intensity variations, points to strong spin–lattice coupling and the possible activation of magnetic correlations under compression. The systematic differences between the Nb- and Ta-based compounds highlight the critical influence of $B$-site–dependent spin–orbit coupling and orbital hybridization on the pressure response of structural and vibrational properties. These results establish pressure as an effective control parameter for tuning lattice, vibrational, and potentially magnetic degrees of freedom in Mn-based $A_4B_2$O$_9$ magnetoelectric systems.

\section{Acknowledgments}
	This work is supported by the National Science Foundation of China (W2432007, 42150101), the National Key Research and Development Program of China (2022YFA1402301), Shanghai Key Laboratory for Novel Extreme Condition Materials, China (No. 22dz2260800), Shanghai Science and Technology Committee, China (No. 22JC1410300). Synchrotron based high-pressure x-ray diffraction experiments have been conducted at Elettra synchrotron, Trieste, Italy through the proposal ID 	20225002; and at Indus-2 Synchrotron, RRCAT, Indore, India through Beamline Reservation Number IBR/3637/2021-11-04/INDUS-2/BL-12 ADXRD. I.K.A. acknowledges the the Indian Institute of Science (IISc), Bengaluru, and the International Centre for Theoretical Physics, Trieste for the IISc-ICTP fellowship. B.J. and I.K.A. acknowledge the "Xpress\_partner" internal project of Elettra-Sincrotrone Trieste.

\clearpage
\setcounter{figure}{0}
\renewcommand{\figurename}{Fig.}
\renewcommand{\thefigure}{S\arabic{figure}}
\setcounter{table}{0}
\renewcommand{\thetable}{S\arabic{table}}

\section{Supplemental material}

\begin{figure}[ht!]
\includegraphics[width = 16cm]{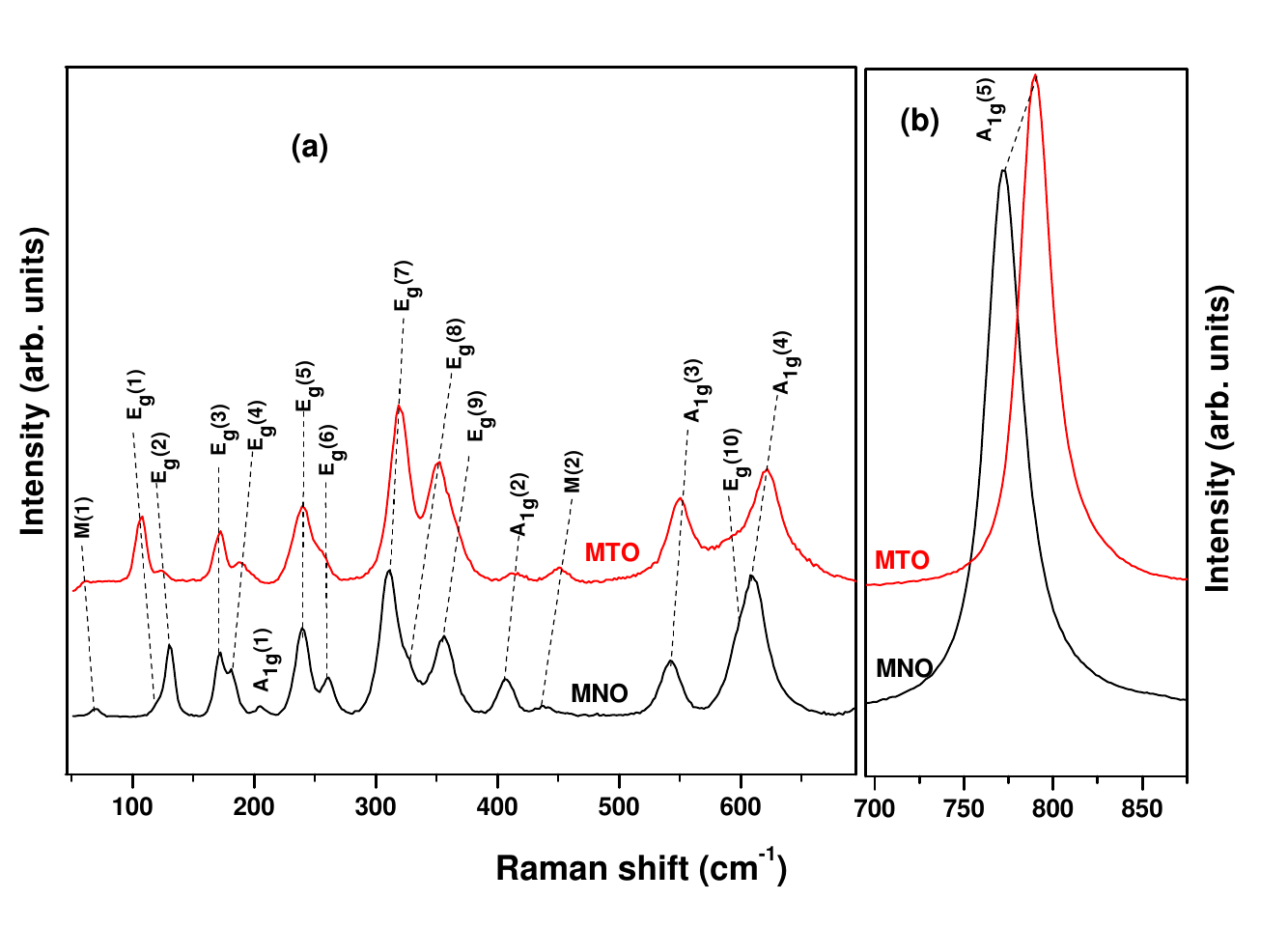}
\caption{\label{fig1} Comparison of Raman spectra of MNO and MTO at Ambient conditions in the frequency range (a) 45-695 cm$^{-1}$; and (b) 695-875 cm$^{-1}$. Mode evolution across two systems are highlighted by the black dashed lines.}
\end{figure}

\begin{figure}[ht!]
\includegraphics[width = 18cm]{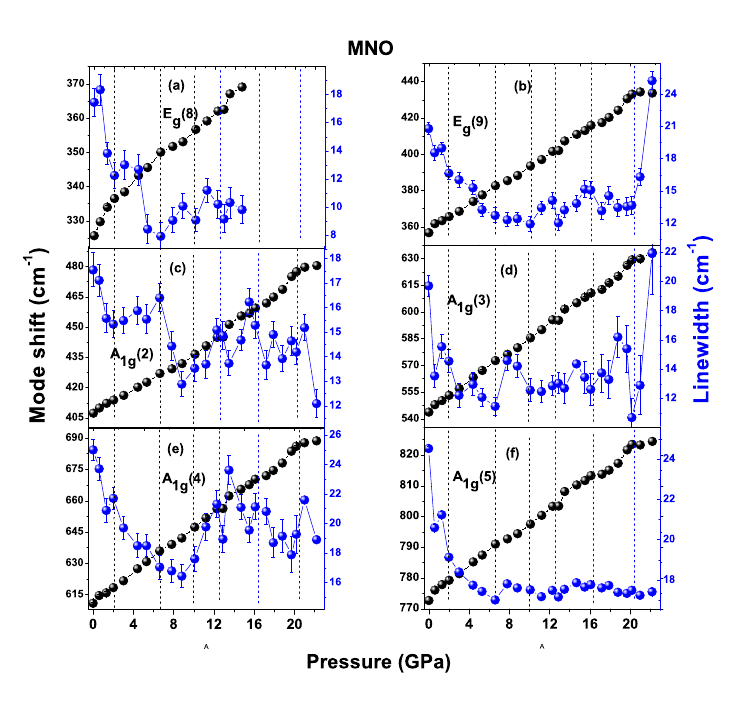}
\caption{\label{fig1} (a)-(f) Pressure evolution of the Raman mode frequencies and linewidths of the high-frequency modes in MNO. Dashed black lines indicate the isostructural transition pressures, while blue dashed lines mark the onset of long-range structural transitions or major structural rearrangements.}
\end{figure}

\begin{figure}
\includegraphics[width=18cm]{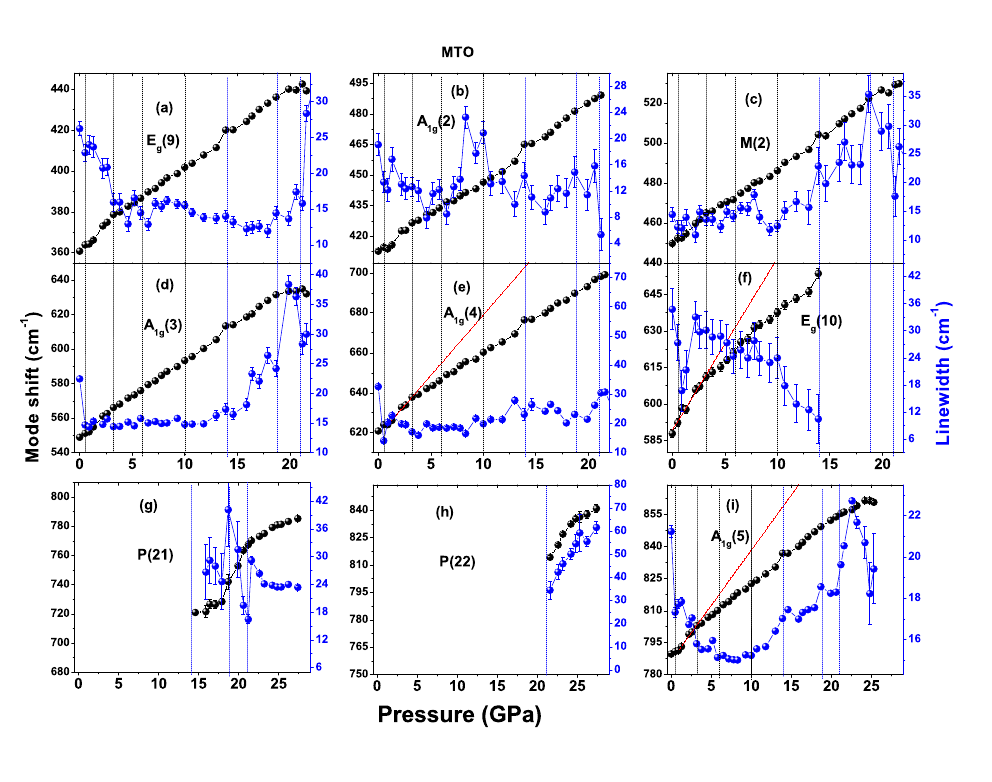}
\caption{\label{fig3} (a)-(i) Pressure dependence of the frequencies and linewidths of the high-frequency Raman modes in MTO. Vertical dashed black lines denote the isostructural transition pressures, whereas blue dashed lines indicate the onset of long-range structural transformations or significant lattice rearrangements. Solid red lines represent the linear fits to the A$_{1g}$(4), E$_{g}$(10), and A$_{1g}$(5) mode frequencies within the 1–3.2 GPa range.}
\end{figure}

\begin{figure}[ht!]
\includegraphics[width=12cm]{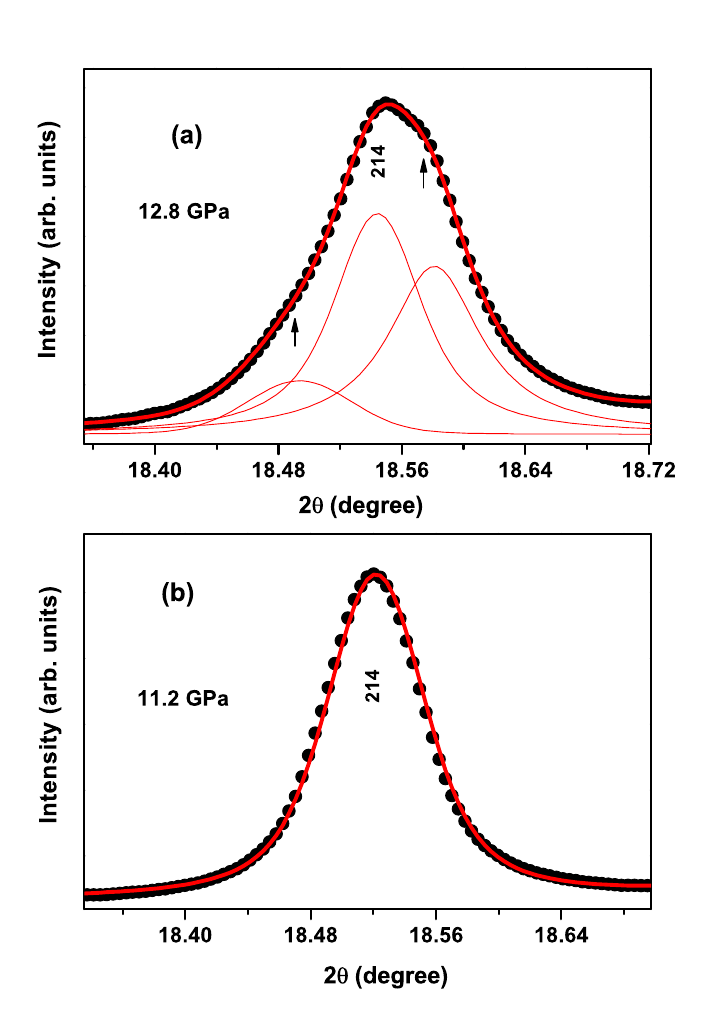}
\caption{\label{fig5} Evolution of the (214) Bragg peak of MNO at (a) 12.8 GPa; and (b) 11.2 GPa. Two black arrows indicate the emergence of additional diffraction peaks. Black circles represent the experimental data, while the red line corresponds to the Voigt-profile fit to the data. }
\end{figure}

\begin{figure*}[ht!]
\includegraphics[width=13cm]{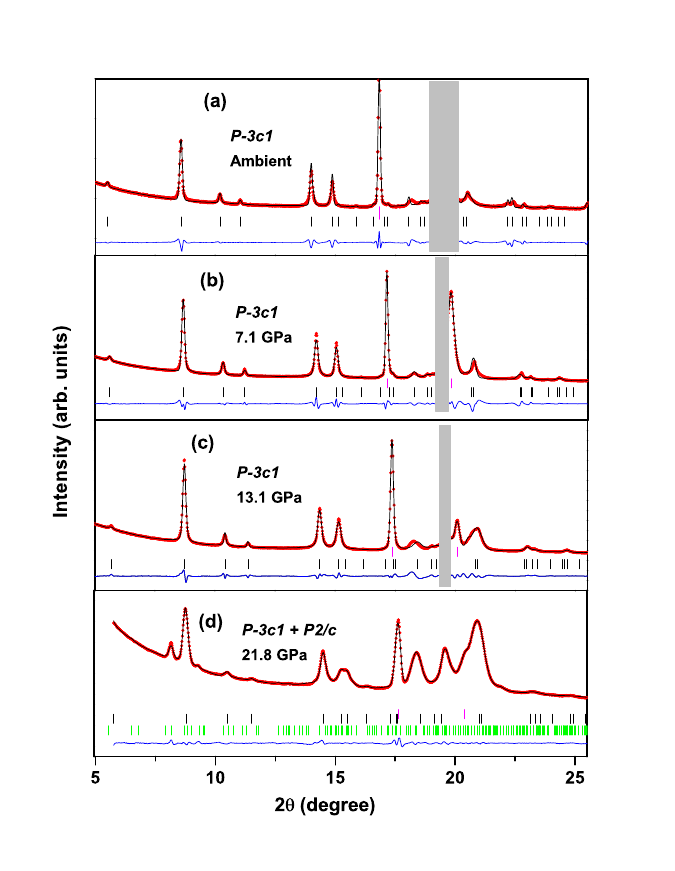}
\caption{\label{fig1} Rietveld-refined XRD patterns of MTO at selected pressures. Panels (a), (b) and (c) correspond to the trigonal \textit{P-3c1} phase at ambient pressure, 7.1 and 13.1 GPa, respectively. Panels (d) shows the coexistence of the \textit{P-3c1} and \textit{P2/c} phases at 21.8 GPa. Red circles represent the observed data, and the solid black line denotes the Rietveld fit. Vertical tick marks indicate the calculated Bragg positions for the \textit{P-3c1} (black) and \textit{P2/c} (green) phases. The blue line at the bottom shows the difference between the experimental and calculated intensities. The contribution from the gasket is masked by the gray shaded region.} 
\end{figure*}

\begin{table}[H]
\caption{\label{tab:tableS1
}
Rietveld refined processed unit cell structural parameters of Mn$_4$Nb$_2$O$_9$ and Mn$_4$Ta$_2$O$_9$ in trigonal \textit{P-3c1} (No. 165, Z = 2) symmetry at ambient pressure.}
\begin{ruledtabular}
\begin{tabular}{ccccccccc}
 Material & a (\AA) & c (\AA) & atom & site & x & y & z & U$_{iso}$ \\
\hline
& & & Mn1 & 4$d$ & 1/3  & 2/3  & 0.0163(7) & 0.0158 (12)\\
& & & Mn2 & 4$d$ & 1/3  & 2/3  & 0.2974(3) & 0.0148(14)\\ 
MNO & 5.3261(5) & 14.3067(4) & Nb &  4$c$ & 0  & 0  & 0.3574(5) & 0.0184(15)\\
& & & O1& 6$f$ & 0.2684(13)  & 0  & 0.25 & 0.0271(17)\\
& & & O2& 12$g$ & 0.3059(11)  & 0.2659(12) &  0.0884(15) & 0.0194(13)\\
\hline
& & & Mn1 & 4$d$ & 1/3  & 2/3  & 0.0184(8) & 0.0176(15)\\
& & & Mn2 & 4$d$ & 1/3  & 2/3  & 0.3051(4) & 0.0151(13)\\ 
MTO & 5.3338(4) & 14.3482(2) & Ta &  4$c$ & 0  & 0  & 0.3566(2) & 0.0163(14)\\
& & & O1& 6$f$ & 0.2865(12)  & 0  & 0.25 & 0.0158(15)\\
& & & O2& 12$g$ & 0.3374(14)  & 0.2932(13) & 0.0941(13) & 0.0178(11)\\

\end{tabular}
\end{ruledtabular}
\end{table}

\begin{table*}[b]
\caption{\label{tab:tableS2
}
Rietveld refined crystal structural parameters of Mn$_4$Nb$_2$O$_9$ at 26.5 GPa in space group \textit{P2/c} (No. 13, Z = 4).}
\begin{ruledtabular}
\begin{tabular}{cccccccccc}
 $a$ (\AA) & $b$ (\AA) & $c$ (\AA)  & $\beta(^\circ)$ & atom & site & x & y & z & U$_{iso}$ \\
\hline
& & & & Mn1 &  4$g$ & 0.4257(4) & 0.1511(5) & 0.9217(2) & 0.0266(11) \\ 
& & & & Mn2 & 2$e$ & 0 & 0.0069(8) & 1/4 & 0.0253(14) \\
& & & & Mn3 & 2$f$ & 1/2 & 0.5267(5) & 1/4 & 0.0249(14) \\
& & & & Mn4 & 4$g$ & 0.0158(8) & 0.6691(5) & 0.9043(4) & 0.0273(13) \\
& & & & Mn5 & 2$a$ & 0 & 0 & 0 & 0.0264(16) \\ 
& & & & Mn6 & 2$b$ & 1/2 & 1/2 & 0 & 0.0281(12) \\
& & & & Nb1 & 4$g$ & 0.50137(3) & 0.16037(6) & 0.16203(6) & 0.0275(14) \\ 
& & & & Nb2 & 4$g$ & 0.0005(9) & 0.6516(4) & 0.1529(7) & 0.0288(16) \\  
5.7624(5) & 8.8689(7) & 12.1898(3) & 91.65(5) & O1 & 4$g$ & 0.7452(12) & 0.2077(14) & 0.2699(13) & 0.0286(15) \\  
& & & & O2 & 4$g$ & 0.1896(13) & 0.7295(13) & 0.2472(14) & 0.0279(16) \\  
& & & & O3 & 2$e$ & 0 & 0.4976(12) & 3/4 & 0.0281(14) \\
& & & & O4 & 2$f$ & 1/2 & 0.0257(15) & 3/4 & 0.0287(17) \\
& & & & O5 & 4$g$ & 0.2914(13) & 0.0751(16) & 0.4230(14) & 0.0291(13) \\
& & & & O6 & 4$g$ & 0.8048(11) & 0.57376(14) & 0.4237(16) & 0.0283(16) \\
& & & & O7 & 4$g$ & 0.0181(16) & 0.1794(15) & 0.8997(13) & 0.0276(15) \\
& & & & O8 & 4$g$ & 0.4695(17) & 0.6829(14) &  0.9222(12) & 0.02751(17) \\
& & & & O9 & 4$g$ & 0.7248(14) &  0.0176(17) & 0.46281(15) & 0.0281(16) \\ 
& & & & O10 & 4$g$ & 0.2531(17) & 0.5701(15) & 0.3968(17) & 0.0285(17) \\
 
\end{tabular}
\end{ruledtabular}
\end{table*}

\newpage  
\begin{table*}[b]
\caption{\label{tab:tableS2
}
Rietveld refined crystal structural parameters of Mn$_4$Ta$_2$O$_9$ at 21.8 in space group \textit{P2/c} (No. 13, Z = 4).}
\begin{ruledtabular}
\begin{tabular}{cccccccccc}
 $a$ (\AA) & $b$ (\AA) & $c$ (\AA)  & $\beta(^\circ$) & atom & site & x & y & z & U$_{iso}$ \\
\hline
& & & & Mn1&  4$g$ & 0.5048(4) & 0.1456(5) & 0.9374(4) & 0.0269(12) \\ 
& & & & Mn2 & 2$e$ & 0 & 0.0159(6) & 1/4 & 0.0273(15) \\
& & & & Mn3 & 2$f$ & 1/2 & 0.5056(7) & 1/4 & 0.0281(14) \\
& & & & Mn4 & 4$g$ & 0.0110(9) & 0.6692(6) & 0.9131(4) & 0.0275(16) \\
& & & & Mn5 & 2$a$ & 0 & 0 & 0 & 0.0279(15) \\ 
& & & & Mn6 & 2$b$ & 1/2 & 1/2 & 0 & 0.0271(14) \\
& & & & Ta1 & 4$g$ & 0.5002(4) & 0.1591(7) & 0.1616(6) & 0.0282(15) \\ 
& & & & Ta2 & 4$g$ & 0.0052(9) & 0.6515(5) & 0.1501(7) & 0.0287(16) \\  
 5.8013(7) & 8.7932(8) & 12.1373(5) & 91.19(6) & O1 & 4$g$ & 0.7407(13) & 0.1931(14) & 0.2778(16) & 0.0278(17) \\  
& & & & O2 & 4$g$ & 0.1515(14) & 0.7728(11) & 0.2519(15) & 0.0289(16) \\  
& & & & O3 & 2$e$ & 0 & 0.5063(12) & 3/4 & 0.0282(15) \\
& & & & O4 & 2$f$ & 1/2 & -0.0076(17) & 3/4 & 0.0276(14) \\
& & & & O5 & 4$g$ & 0.2869(15) & 0.0817(18) & 0.4091(15) & 0.0284(17) \\
& & & & O6 & 4$g$ & 0.8241(13) & 0.5561(14) & 0.4308(16) & 0.0287(16) \\
& & & & O7 & 4$g$ & 0.0384(17) & 0.1724(15) & 0.9208(11) & 0.0277(15) \\
& & & & O8 & 4$g$ & 0.4597(16) & 0.6771(16) &  0.9190(13) & 0.0289(17) \\
& & & & O9 & 4$g$ & 0.6835(14) &  0.0416(17) & 0.4711(15) & 0.0283(15) \\ 
& & & & O10 & 4$g$ & 0.2529(15) & 0.5599(14) & 0.3979(18) & 0.0281(16) \\
 
\end{tabular}
\end{ruledtabular}
\end{table*}

\end{document}